\documentclass[iop]{emulateapj}
\usepackage{amsmath,amssymb}
\usepackage{xspace}
\usepackage{graphicx}
\usepackage{epstopdf}
\usepackage{graphicx}
\usepackage{epsfig}

\def\chandra{{\it Chandra\/}}
\def\xray{\hbox{X-ray}}

\def\lum{erg~s$^{-1}$}
\def\flux{erg~cm$^{-2}$~s$^{-1}$}

\def\lsim{\mathrel{\rlap{\lower4pt\hbox{\hskip1pt$\sim$}}
    \raise1pt\hbox{$<$}}}                
\def\gsim{\mathrel{\rlap{\lower4pt\hbox{\hskip1pt$\sim$}}
    \raise1pt\hbox{$>$}}}                
\def\cdfs{\hbox{CDF-S}}

\newcommand{\degree}{\ensuremath{^\circ}}

\slugcomment{}
\shorttitle{Source Population Responsible for Unresolved 6--8~keV XRB}
\shortauthors{Xue et~al.}

\begin{document}

\title{Tracking down the source population responsible for the unresolved cosmic 6--8~keV background}


\author{Y. Q. Xue\altaffilmark{1,2,3}, 
S. X. Wang\altaffilmark{1,2}, 
W. N. Brandt\altaffilmark{1,2},
B. Luo\altaffilmark{1,2},
D.~M.~Alexander\altaffilmark{4},
F.~E.~Bauer\altaffilmark{5,6},
A.~Comastri\altaffilmark{7},
A.~C.~Fabian\altaffilmark{8},
R.~Gilli\altaffilmark{7},
B.~D.~Lehmer\altaffilmark{9,10},
D.~P.~Schneider\altaffilmark{1,2},
C.~Vignali\altaffilmark{11}, and
M.~Young\altaffilmark{1,2}
}

\altaffiltext{1}{Department of Astronomy and Astrophysics, Pennsylvania State University, University Park, PA 16802, USA}
\altaffiltext{2}{Institute for Gravitation and the Cosmos, Pennsylvania State University, University Park, PA 16802, USA}
\altaffiltext{3}{Key Laboratory for Research in Galaxies and Cosmology, Department of Astronomy, University of Science and Technology of China, Chinese Academy of Sciences, Hefei, Anhui 230026, China; xuey@ustc.edu.cn}
\altaffiltext{4}{Department of Physics, Durham University, Durham, DH1 3LE, UK}
\altaffiltext{5}{Pontificia Universidad Cat\'{o}lica de Chile, Departamento de Astronom\'{\i}a y Astrof\'{\i}sica, Casilla 306, Santiago 22, Chile}
\altaffiltext{6}{Space Science Institute, 4750 Walnut Street, Suite 205, Boulder, CO 80301, USA}
\altaffiltext{7}{INAF---Osservatorio Astronomico di Bologna, Via Ranzani 1, Bologna, Italy}
\altaffiltext{8}{Institute of Astronomy, Madingley Road, Cambridge, CB3 0HA, UK}
\altaffiltext{9}{The Johns Hopkins University, Homewood Campus, Baltimore, MD 21218, USA}
\altaffiltext{10}{NASA Goddard Space Flight Centre, Code 662, Greenbelt, MD 20771, USA}
\altaffiltext{11}{Universit\'a di Bologna, Via Ranzani 1, Bologna, Italy}




\begin{abstract}

Using the 4~Ms \chandra\ Deep Field-South (CDF-S) survey, we have
identified a sample of 6845 \xray\ undetected galaxies that dominates
the unresolved \hbox{$\approx 20$--25\%} of the 6--8~keV cosmic \xray\ background
(XRB).
This sample was constructed by applying mass and color cuts to sources 
from a parent catalog based on GOODS-South {\it HST}
$z$-band imaging of the central 6\arcmin-radius area of the 4~Ms CDF-S.
The stacked 6--8~keV detection is significant at the $3.9\sigma$ level,
but the stacked emission was not detected in the 4--6~keV band
which indicates the existence of an underlying population of highly obscured
active galactic nuclei (AGNs).
Further examinations of these 6845 galaxies indicate that 
the galaxies on the top of the blue cloud and
with redshifts of $1\lsim z\lsim 3$, magnitudes of \hbox{$25\lsim z_{\rm 850}\lsim 28$}, 
and stellar masses of $2\times 10^8\lsim M_\star/M_\odot\lsim 2\times 10^9$ 
make the majority contributions to the unresolved \hbox{6--8~keV} XRB.
Such a population is seemingly surprising 
given that the majority of the \xray\ detected AGNs reside in 
massive ($\gsim 10^{10}\:M_\odot$) galaxies.
We discuss constraints upon this underlying AGN population, 
supporting evidence for relatively low-mass galaxies hosting highly obscured AGNs,
and prospects for further boosting the stacked signal.

\end{abstract}


\section{Introduction}\label{sec:intro}

Deep extragalactic \xray\ surveys 
have been effective in finding active galactic nuclei (AGNs) out to $z\approx 5$
(see, e.g., Brandt \& Hasinger 2005 and Brandt \& Alexander 2010 for reviews),
with an observed AGN sky density approaching 14,900 deg$^{-2}$
achieved by the deepest \xray\ surveys, the \chandra\ Deep Fields 
(e.g., Brandt et~al. 2001; Giacconi et~al. 2002; Alexander et~al. 2003; Bauer et~al. 2004;
Luo et~al. 2008; Xue et al. 2011, hereafter X11; Lehmer et al.~2012).
In both broad (\hbox{0.5--2} and \hbox{2--8 keV}) and
narrow (\hbox{0.5--1}, \hbox{1--2}, \hbox{2--4}, \hbox{4--6}, and \hbox{6--8 keV}) energy bands,
\hbox{$\approx 75$--95\%} of the cosmic \xray\ background (XRB) 
emission has been resolved into discrete sources
(e.g., Brandt et~al. 2000; Mushotzky et~al. 2000;
Bauer et~al. 2004; Worsley et~al. 2005; Hickox \& Markevitch 2006; Luo et~al. 2011), 
the majority of which are moderately to highly obscured AGNs
(e.g., Barger et~al. 2003; Szokoly et~al. 2004; Tozzi et~al. 2006).
Of particular interest is the remaining unresolved XRB at the highest energies 
accessible to \chandra, \hbox{6--8~keV}, since XRB synthesis models
(e.g., Gilli et~al. 2007) indicate that much of this emission should arise from
the highly obscured AGNs that contribute strongly to the XRB near its \hbox{$\approx20$--40~keV} peak.

Recently, utilizing the 4~Ms \chandra\ Deep Field-South (CDF-S; X11) data,
Luo et~al. (2011) found that
the resolved \hbox{6--8 keV} XRB fraction is $\approx 78\% \pm 6\%$,
taking into account both the \xray\ source contribution and bright-end correction,
and adopting the XRB normalization from Hickox \& Markevitch
(2006; the XRB has a power-law spectral slope of $\Gamma=1.4$ and a normalization
of 10.9 photons s$^{-1}$ keV$^{-1}$ sr$^{-1}$ 
at 1~keV).\footnote{Throughout this paper, 
the resolved \hbox{6--8 keV} XRB fraction 
refers to the ratio between the detected \hbox{6--8 keV}
surface brightness in the field and the \hbox{6--8 keV} XRB level
determined by Hickox \& Markevitch (2006).}
Luo et~al. (2011) further found that 
the unresolved \hbox{$\approx 20$--25\%} of the XRB
in the \hbox{6--8 keV} band 
can plausibly be explained by the stacked emission (a $2.5\sigma$ signal)
from a sample of 18,272 \xray\ undetected \hbox{GOODS-South} (\hbox{GOODS-S})
{\it HST} \hbox{$z$-band} sources.
The above resolved fraction should be considered cautiously
as it is known that there is cosmic variance,
likely arising from the underlying large-scale structure 
variations between fields, in the deepest \chandra\ surveys 
(e.g., Barger et~al. 2002, 2003; Cowie et~al. 2002;  
Gilli et~al. 2003, 2005; Yang et~al. 2003;
Hickox \& Markevitch 2006; Silverman et~al. 2010).
The resultant uncertainty in the XRB normalization is likely in the
range \hbox{10--20\%} (e.g., Hickox \& Markevitch 2006).
In this field Luo et~al. (2011) obtained
the above \hbox{6--8 keV} stacked signal
and it is of interest to understand its origin.
Luo et~al. (2011) 
also showed that there should be an underlying population of
highly obscured AGNs hidden among the aforementioned 
\xray\ undetected galaxies.
Thanks to superb sensitivities, 
ultradeep \xray\ observations have already been able to
reveal a significant fraction of such previously ``missing'' 
highly obscured AGNs (e.g., Alexander et~al. 2011; Comastri et~al. 2011; 
Feruglio et~al. 2011; Gilli et~al. 2011; X11)
that are estimated to be roughly as numerous as moderately obscured AGNs (e.g., Gilli et~al. 2007).

Mounting evidence has shown that luminous AGNs tend to reside in massive 
(i.e., $M_\star\gsim 10^{10}\:M_\odot$; $M_\star$ is stellar mass) and red galaxies
from the local universe up to \hbox{$z\approx 3$--4} 
(e.g., Barger et~al. 2003; Bundy et~al. 2008; Brusa et~al. 2009; Silverman et~al. 2009; Xue et~al. 2010, hereafter X10; Mullaney et~al. 2012).
In this paper, we thus focus on using these mass and color constraints as clues to hunt for
an underlying population of highly obscured AGNs responsible
for the unresolved \hbox{$\approx 20$--25\%} of the \hbox{6--8 keV} XRB.
This paper is structured as follows:
in \S~\ref{sec:prop} we describe how we estimated physical properties
for sources of interest;
in \S~\ref{sec:results} we present the source-stacking analysis
and the results obtained;
in \S~\ref{sec:robustness} we assess the robustness of the stacking results;
and in \S~\ref{sec:discussion} we discuss the implications of
the results.

Throughout, a cosmology of $H_0=70.4$~km~s$^{-1}$~Mpc$^{-1}$, $\Omega_m=0.272$, and $\Omega_{\Lambda}=0.728$ is adopted (e.g., Komatsu et~al. 2011).
Unless stated otherwise, apparent magnitudes are given in the AB system (Oke \& Gunn 1983), 
and rest-frame absolute magnitudes are given in the Vega system (Johnson \& Morgan 1953).
We adopt a Galactic column density of $N_{\rm H}=8.8\times 10^{19}\:{\rm cm}^{-2}$ 
(e.g., Stark et al. 1992) along the line of sight to the \cdfs\ and correct for 
Galactic extinction in all relevant \xray\ analyses below.

\section{Source Properties}\label{sec:prop}

\begin{figure*}[th]
\center
\includegraphics[angle=0,scale=.75]{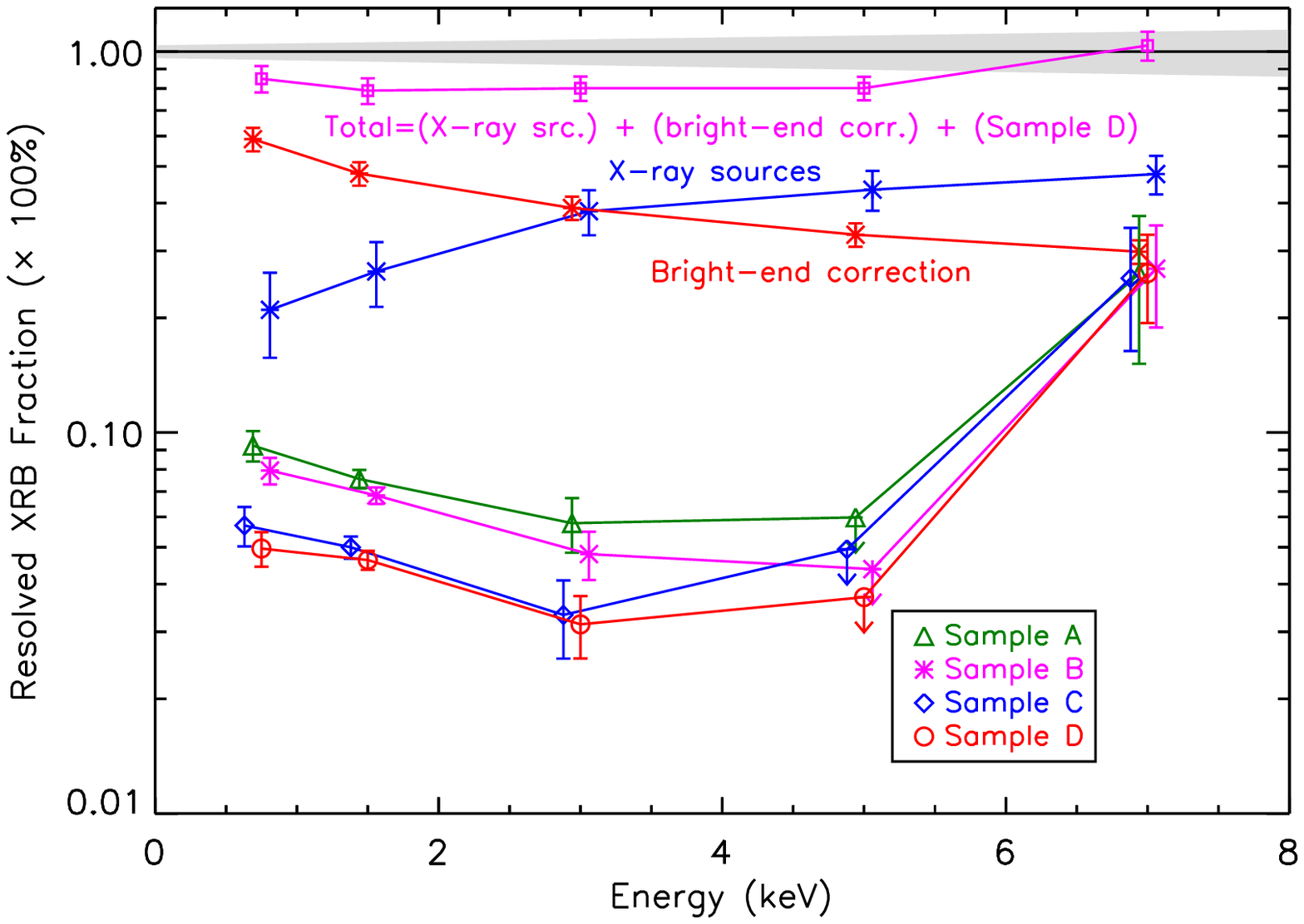}
\caption{Resolved XRB fractions in five energy bands between 0.5 and 8~keV;
this figure illustrates the values in Table~\ref{tab:resolved}
(details are therefore referred to Table~\ref{tab:resolved};
cf. Fig.~6 of Luo et~al. 2011).
The data points are shifted slightly in the $x$-direction for clarity.
The total XRB intensities 
are adopted from Hickox \& Markevitch (2006; also see \S~\ref{sec:intro})
with uncertainties indicated by the gray area.
}
\label{fig:xrbfrac}
\end{figure*}

In this section we describe briefly how we estimated source properties, i.e., 
redshifts, rest-frame absolute magnitudes, and stellar masses, 
for sources of interest.
The Luo et~al. (2011) sample of 18,272 
\xray\ undetected \hbox{GOODS-S} {\it HST} \hbox{$z$-band} sources 
is located within 6\arcmin\ of the 4~Ms \cdfs\ average aim point
($\alpha_{\mbox{J2000.0}} = 03^{\rm h} 32^{\rm m} 28.06^{\rm s} $,
$\delta_{\mbox{J2000.0}} = -27^{\circ} 48\arcmin 26.4\arcsec$)
and outside of twice the 90\% encircled-energy (in the \hbox{0.5--2 keV} band) aperture radius
of any 4~Ms \cdfs\ main-catalog source\footnote{As described in X11, 
the 4~Ms \cdfs\ main catalog contains 740 \xray\ sources 
that are detected with {\sc wavdetect} at a false-positive probability threshold 
of $10^{-5}$ in at least one of three \xray\ bands (0.5--8~keV, full band; 0.5--2~keV, soft band; 
and 2--8~keV, hard band) and also satisfy a binomial-probability source-selection criterion 
of $P < 0.004$ (i.e., the probability of sources not being real is less than 0.004).
The flux limits at an off-axis angle of 6\arcmin\ for the 4~Ms \cdfs\ are
$\approx 1.2\times 10^{-16}$, $3.1\times 10^{-17}$, and $2.6\times 10^{-16}$~\flux\
for the full, soft, and hard bands, respectively.}
(the resultant total stacking area is 0.027 deg$^2$).
As shown in Luo et~al. (2011), this sample appears to be
responsible for the unresolved \hbox{$\approx 20$--25\%} of the \hbox{6--8 keV} XRB.
Discarding the 160 sources with a low signal-to-noise ratio 
that were not included in the Dahlen et~al. (2010) catalog (detailed below) 
and the 77~stars that were spectroscopically identified therein 
(see \S~2.3 of Dahlen et~al. 2010 for the references of the spectroscopic data used),
we reduce the size of the above Luo et~al. (2011) sample to 18,035 and refer to this reduced sample
as ``Sample~A'' hereafter.
The properties and contribution to the \hbox{6--8 keV} XRB for Sample~A are listed in 
Table~\ref{tab:samples},
which shows that
Sample~A still appears to be responsible for the unresolved 
\hbox{6--8 keV} signal seen by Luo et~al. (2011;
see \S~\ref{sec:results} for the details of the stacking procedure).
We provide in Table~\ref{tab:resolved} resolved XRB fractions in
various bands between 0.5 and 8~keV for Sample~A and additional samples 
of interest (see \S~\ref{sec:results}). 
We also directly illustrate the values
in Table~\ref{tab:resolved}
as Fig.~\ref{fig:xrbfrac}.

It can be inferred from
Fig.~\ref{fig:xrbfrac} (i.e., the top-most summed data points
shown as squares) that there should be
a yet-to-be-identified source population
that accounts for the remaining $<6$~keV
emission without boosting significantly
the \hbox{6--8 keV} emission.
As shown in \S~\ref{sec:results},
this missing source population cannot be associated with individual
galaxies, which would otherwise have been stacked already.
It is likely that this remaining $<6$~keV emission is from
groups/clusters (e.g., Cappelluti et~al. 2012), 
whose emission would not be included in our
stacking of galaxies and whose spectrum often has a strong thermal cutoff
thus contributing emission in the soft band 
but not much in the hard band.
It is also likely that cosmic variance might play some role here
(e.g., affecting the shape of the summed spectrum 
shown in Fig.~\ref{fig:xrbfrac}).
Given the complexities in determining resolved XRB fractions
(e.g., adopting various methodologies and different XRB normalizations),
the resolved XRB fractions reported in Table~\ref{tab:resolved} (i.e., summed
contributions of \xray\ sources and bright-end correction
that range from $\approx 75\%$ to $80\%$)
are in general agreement with those from other works.
For example, Hickox \& Markevitch (2006) found 
the resolved XRB fractions to be $77\pm3\%$ and $80\pm8\%$
for the \hbox{1--2} and \hbox{2--8 keV} bands, respectively;
Lehmer et~al. (2012) obtained resolved XRB fractions of 
$76\pm4\%$, $82\pm13\%$, $88\pm14\%$, 
and $82\pm9\%$ for the \hbox{0.5--2}, \hbox{2--8}, \hbox{4--8}, 
and \hbox{0.5--8 keV} bands, respectively.

\begin{table*}
\caption{Stacked 6--8 keV Properties}
\resizebox{\textwidth}{!}{%
\begin{tabular}{ccrccccccccc}\hline\hline
 {} &
 {} &
 {} &
 {Median} &
 {Median} &
 {Median} &
 {Net} &
 {S/N} &
 {Band} &
 {} &
 {Total} &
 {Resolved} \\
 {Sample} &
 {Criteria} &
 {$N_{\rm gal}$} &
 {$z$} &
 {$M_\star$ ($M_\odot$)}&
 {$C_{\rm eff}$} &
 {Counts} &
 {($\sigma$)} &
 {Ratio} &
 {$\Gamma_{\rm eff}$} &
 {Intensity} &
 {Fraction (\%)} \\
 {(1)} &
 {(2)} &
 {(3)} &
 {(4)} &
 {(5)} &
 {(6)} &
 {(7)} &
 {(8)} &
 {(9)} &
 {(10)} &
 {(11)} &
 {(12)} \\ \hline
A & All galaxies within 6\arcmin\ of aim point &    18,035   &    1.10 & $2.3\times 10^8$ &  $-$0.53 &  819$\pm$343 &  2.4 &  0.44$\pm$0.08  &  1.67$\pm$0.16 & 1.27  &  26.1$\pm$10.9 \\
B & $M_\star\ge 2\times 10^8\:M_\odot$ & 9542  &  1.48 &  $8.9\times 10^8$ & $-$0.45  & 841$\pm$250  &  3.4  & 0.46$\pm$0.07  &     1.63$\pm$0.13 &   1.32 &  26.9$\pm$8.0 \\
C & $-0.85<C_{\rm eff}<0$  &  12,290  &  1.13 & $2.6\times 10^8$  &   $-$0.51  &  799$\pm$283   &  2.8  & 0.47$\pm$0.10  & 1.62$\pm$0.20  &  1.24  &  25.4$\pm$9.0 \\
D & $M_\star\ge 2\times 10^8\:M_\odot$ \& $-0.85<C_{\rm eff}<0$ &  6845  & 1.59 &  $8.1\times 10^8$ &  $-$0.48 &  820$\pm$212  &  3.9 & $0.48\pm0.08$  &  $1.60\pm0.16$ &  1.28 &  26.2$\pm$6.8 
\\\hline
\end{tabular}}
{\sc Note.} --- 
Col.~(1): Sample of galaxies used for stacking. Samples B--D are subsets of Sample~A.
Col.~(2): Criteria used to define the stacked sample (see \S~\ref{sec:results} for the definition of $C_{\rm eff}$).
Col.~(3): Number of galaxies in the stacked sample.
Col.~(4): Median redshift of the stacked sample.   
Col.~(5): Median stellar mass of the stacked sample.   
Col.~(6): Median effective rest-frame color of the stacked sample.   
Col.~(7): Stacked net source counts in the 6--8~keV band, with $1\sigma$ Gaussian statistical errors.  
Col.~(8): Stacked signal-to-noise ratio in the 6--8~keV band.   
Col.~(9): Stacked band ratio, defined as the ratio between the observed 2--8~keV and 0.5--2~keV count rates. The $1\sigma$ errors were calculated following the ``numerical method'' described in \S1.7.3 of Lyons (1991).   
Col.~(10): Effective photon index with $1\sigma$ errors of the stacked sample.   
Col.~(11): Total 6--8~keV intensity (in units of $10^{-12}$~\flux~deg$^{-2}$) of the stacked sample.   
We calculated effective photon indices and fluxes based on band ratios and aperture-corrected
count rates using the CXC's Portable Interactive Multi-Mission Simulator.
Col.~(12): Resolved fraction of the 6--8~keV XRB from the stacked sample.
We adopted the XRB normalization from Hickox \& Markevitch (2006); see \S~\ref{sec:intro}.
\label{tab:samples}
\end{table*}

\begin{table*}
\caption{Resolved XRB Fractions in Various Bands}
\begin{tabular}{cccccc}\hline\hline
  & 0.5--1 keV & 1--2 keV & 2--4 keV & 4--6 keV & 6--8 keV \\ 
Sample & Resolved Frac. (\%) & Resolved Frac. (\%) & Resolved Frac. (\%) & Resolved Frac. (\%) & Resolved Frac. (\%) \\
(1) & (2) & (3) & (4) & (5) & (6) \\\hline
A & $9.2\pm0.8$ & $7.6\pm0.4$ & $5.8\pm0.9$ & $<6.0$ & $26.1\pm10.9$ \\
B & $7.9\pm0.6$ & $6.8\pm0.3$ & $4.8\pm0.7$ & $<4.4$ & $26.9\pm8.0$ \\
C & $5.7\pm0.7$ & $5.0\pm0.3$ & $3.3\pm0.8$ & $<4.9$ & $25.4\pm9.0$ \\
D & $5.0\pm0.5$ & $4.6\pm0.3$ & $3.1\pm0.6$ & $<3.7$ & $26.2\pm6.8$ \\
X-ray sources & $21.0\pm5.3$ & $26.5\pm5.1$ & $38.1\pm5.2$ & $43.4\pm5.2$ & $47.7\pm5.5$ \\
Bright-end correction & $58.9\pm4.2$ & $47.8\pm3.4$ & $38.9\pm2.7$ & $33.1\pm2.3$ & $29.8\pm2.1$ \\
X-ray src. + corr. + D & $84.8\pm6.7$ & $78.9\pm6.1$ & $80.1\pm5.9$ & $76.5_{-5.7}^{+6.8}$ & $103.7\pm9.1$ \\
\hline
\end{tabular}
\\Notes. --- 
Col. (1):
Samples~A, B, C, and D are the same as those in Table~\ref{tab:samples}.
The fifth and sixth rows represent XRB fractions resolved by 
\xray\ point sources in the 4~Ms \cdfs\ and the corresponding bright-end
correction (Luo et~al. 2011; also see \S~\ref{sec:intro}).
The last row is the sum of contributions from \xray\ point sources,
the bright-end correction, and Sample~D.
Cols. (2--6): Resolved XRB fractions and $1\sigma$ uncertainties
in various bands.
In Col. (5), the ``$<$'' signs 
for Samples~A, B, C, and D indicate $3\sigma$ upper
limits on resolved fractions in the \hbox{4--6 keV} band, where
the upper limit for Sample~D was used to determine the upper error
of the total contribution in this band.
We adopted the XRB normalization from Hickox \& Markevitch (2006); see \S~\ref{sec:intro}.
\label{tab:resolved}
\end{table*}

\subsection{Redshifts}\label{sec:z}

Sample~A has a $5\sigma$ $z$-band limiting magnitude of 28.1,
much deeper than most of the photometric-redshift catalogs available in this field
(e.g., Cardamone et~al. 2010; Rafferty et~al. 2011),
and thus $\approx 40\%$ of the Sample~A sources (most with $z_{\rm 850}>26$) have no
photometric-redshift estimates in those catalogs. 
Recently, Dahlen et~al. (2010) derived photometric redshifts ($z_{\rm phot}$'s)
for the 32,508 GOODS-S $z$-band sources in the entire GOODS-S region, 
which include all the 18,035 Sample~A sources.
We chose to re-derive $z_{\rm phot}$'s for these 32,508 sources
in order to ensure consistency of our analyses here 
(i.e., using an optimized comprehensive set of spectral energy distribution
templates throughout; see below) and also include the latest
CANDELS {\it HST}/WFC3 photometry.

We used the ultradeep 12-band photometry and 1382 secure spectroscopic redshifts ($z_{\rm spec}$'s)
assembled by Dahlen et~al. (2010; see their \S~2.3 for how the
quality flag for $z_{\rm spec}$ was assigned).
The 12-band photometry, covering a wavelength range of \hbox{$\approx0.3$--8.0 $\mu$m}
in the observed frame,
consists of the VLT/VIMOS \hbox{$U$-band};
{\it HST}/ACS F435W, F606W, F775W, and F850LP bands;
VLT/ISAAC $J$, $H$, and $K_s$ bands; and {\it Spitzer}/IRAC 3.6, 4.5, 5.8, and 8.0$\mu$m bands.
The photometry (including upper limits\footnote{For photometry
reported by TFIT with a negative flux value
or a positive flux value that is equal or smaller than its error $\sigma_{\rm f}$
(i.e., with a $\le 1$ signal-to-noise ratio), we incorporate this
information as a flux upper limit (i.e., with coverage but no detection) 
by setting both the values of flux and its error to be $\sigma_{\rm f}$.\label{ft:detection}})
was obtained using the TFIT algorithm that 
performs point-spread-function matched photometry
uniformly across different instruments and filters
(see Dahlen et~al. 2010 for details).
Additionally, we also included the latest photometry in the {\it HST}/WFC3 F105W ($Y$),
F125W ($J$), and F160W ($H$) bands, based on the first 10-epoch 
\hbox{GOODS-S} images from CANDELS (Grogin
et~al. 2011; Koekemoer et~al. 2011) that are publicly available. 
For the CANDELS Deep and Wide regions that have multi-epoch $J$ and $H$
coverage, the images were stacked for each band in each region using
the published weight maps by the CANDELS group.
Object catalogs were generated using SExtractor (Bertin \& Arnouts
1996) version 2.8.6, and then source matching was
performed with the Dahlen et~al.~(2010) catalog using the SExtractor
ASSOC option, searching for the nearest match within 0.5\arcsec.
The $J$-band photometry was done with the SExtractor single-image mode,
while the $H$-band photometry was extracted with the dual-image mode to
match detections in $J$. The photometry on $Y$-band images was extracted
only using the single-image mode because of its different sky coverage
from $J$ and $H$ images.
For sources that have $J$ and $H$ detections in both the {\it HST}/WFC3 and VLT/ISAAC,
we found the two sets of $J$ and $H$ measurements in reasonable agreement
without any apparent systematic offsets;
for these sources, we chose to adopt the measurements from {\it HST}/WFC3 and discard 
those from VLT/ISAAC because the {\it HST}/WFC3 photometry is of higher quality.
The $5\sigma$ limiting magnitudes for point sources are the following:
28.0 for the VIMOS \hbox{$U$-band}; 
28.7, 28.8, 28.3, and 28.1 for the ACS 
F435W, F606W, F775W, and F850LP bands;
25.0, 24.5, and 24.4 for the ISAAC $J$, $H$, and $K_s$ bands;
26.1, 25.5, 23.5, and 23.4 for the IRAC 3.6, 4.5, 5.8, and 8.0$\mu$m bands
(see \S~2.1 of Dahlen et~al. 2010 for details);
and 28.2, 27.9, and 27.6 for the WFC3 F105W, F125W, and F160W bands
(see Table~6 of Grogin et~al. 2011 for details),
respectively.

Utilizing the Zurich Extragalactic Bayesian Redshift Analyzer (ZEBRA; Feldmann et~al. 2006),
we adopted a procedure similar to that detailed in Luo et~al. (2010) and Rafferty
et~al. (2011) to calculate $z_{\rm phot}$'s down to $z_{\rm 850}\approx28.1$.
As is standard practice, we constructed
our galaxy spectral energy distribution (SED) templates 
based on the stellar population synthesis model by Bruzual \& Charlot
(2003) with a Chabrier initial-mass function (IMF; Chabrier 2003) and a dust-extinction law
from Calzetti et~al.~(2000). The adopted star-formation history is of
exponential form, $e^{-t/\tau}$, with $\log({\tau/\mbox{year})}$
ranging from 6.5 to 11.0 and $\log({\mbox{age/year})}$ ranging from 
7.0 to 10.1 (both in steps of~0.1). The dust extinction $A_V$ varies between $0$ and $3.0$
(also in steps of~0.1), and the
metallicities are $Z = 0.004,\ 0.008,\ 0.02$ (roughly solar), and 0.05.
Using the available 1382 secure $z_{\rm spec}$'s (redshifts were fixed to the $z_{\rm spec}$ values
for training purposes),
we first ran ZEBRA to identify and apply systematic offsets in the photometry
(differing from filter to filter; typically $\lsim0.3$~mag)
that minimized the residuals between observed and best-fit template fluxes.
We then used ZEBRA to construct new templates by modifying the original templates 
based on the best fits between the corrected photometry and original templates.
Finally we ran ZEBRA on all 32,508 sources, using the corrected photometry and
an improved set of templates, to derive $z_{\rm phot}$'s.

\begin{figure*}[th]
\center
\includegraphics[angle=0,scale=.60]{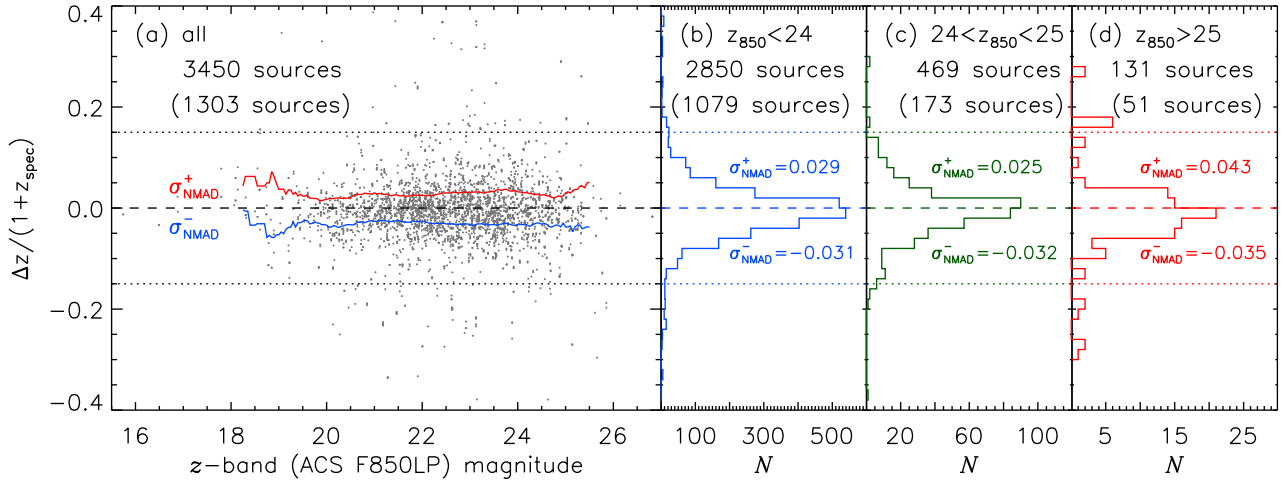}
\caption{Blind-test results of photometric redshifts for all sources with $z_{\rm spec}$
(i.e., including duplicate ones). (a) $\Delta z/(1+z_{\rm spec})$ as a function of $z$-band magnitude.
The $\sigma^+_{\rm NMAD}$ and $\sigma^-_{\rm NMAD}$ running curves (both computed
in bins of $\Delta z_{\rm 850}=1$~mag) are shown as red and blue
curves, respectively.
(b--d) Histograms of $\Delta z/(1+z_{\rm spec})$ in various intervals of $z_{\rm 850}$ magnitude, 
with corresponding values of 
$\sigma^+_{\rm NMAD}$ and $\sigma^-_{\rm NMAD}$ annotated.
In each of the four panels (a--d), the number of all sources is shown
without parentheses and the number of unique sources (i.e., excluding duplicate ones) is shown in
parentheses. The dashed line indicates $\Delta z/(1+z_{\rm spec})=0$, and the dotted lines 
indicate the threshold values of outliers [i.e., $\Delta z/(1+z_{\rm spec})=\pm 0.15$].
}
\label{fig:zphot}
\end{figure*}

We used the normalized median absolute deviation
\begin{equation} 
\sigma_{\rm NMAD}=1.48\times {\rm median}\left(\left|\frac{\Delta
  z-{\rm median}(\Delta z)}{1+z_{\rm
    spec}}\right|\right),\label{equ:nmad}
\end{equation}
where $\Delta z=z_{\rm phot}-z_{\rm spec}$, and the outlier fraction
[outliers are defined as sources with $|\Delta z/(1+z_{\rm spec})|>0.15$]
to assess the $z_{\rm phot}$ quality.
For the spectroscopic subsample, we find 
$\sigma_{\rm NMAD}=0.005$ and an outlier fraction of 1.8\%.
However, the above evaluation cannot represent the real quality of the $z_{\rm phot}$'s,
because the SED templates were modified using the $z_{\rm spec}$ information, 
and we are thus biased to get optimal fitting results for the spectroscopic subsample.
Therefore, to obtain a realistic estimate of the $z_{\rm phot}$ quality for 
the sources lacking $z_{\rm spec}$'s (i.e., no training possible),
we performed a series of blind tests.
We randomly selected 3/4 of the $z_{\rm spec}$ sources
to go through the above training procedure (i.e., photometry correction and
template improvement).
We then derived $z_{\rm phot}$'s for the remaining 1/4 $z_{\rm spec}$ sources
(i.e., blind-test sources whose $z_{\rm spec}$ information was not utilized)
based on the corrected photometry and the expanded set of templates.
The blind test was repeated ten times to ensure a statistically meaningful assessment,
which means that there are duplicated blind-test sources because
a $z_{\rm spec}$ source will on average be used for blind testing 2.5~times.
Figure~\ref{fig:zphot} shows the $z_{\rm phot}$ quality results from the blind tests.
We obtained $\sigma_{\rm NMAD}=0.043$ and an outlier fraction of 7.1\%
for the blind tests.

We defined $\sigma^+_{\rm NMAD}$ and $\sigma^-_{\rm NMAD}$ to
examine further the $z_{\rm phot}$ accuracy as a function of $z_{\rm 850}$ magnitude,
where $\sigma^+_{\rm NMAD}$ is calculated for sources with $z_{\rm phot}>z_{\rm spec}$ using Eq.~\ref{equ:nmad}
and $\sigma^-_{\rm NMAD}$ for sources with $z_{\rm phot}<z_{\rm spec}$. 
The $\sigma^+_{\rm NMAD}$ and $\sigma^-_{\rm NMAD}$ running curves are shown in 
Fig.~\ref{fig:zphot}a (red and blue curves),
both of which are roughly constant ($\approx \pm 0.03$) and symmetric around the $\Delta z/(1+z_{\rm spec})=0$ axis
(dashed line) across a wide range of $z_{\rm 850}$ magnitude.
Therefore, our $z_{\rm phot}$ quality appears to be reasonably accurate
and free of strong systematics down to faint magnitudes, as
can also be inferred from \hbox{Figs.~\ref{fig:zphot}b--\ref{fig:zphot}d} that show the
histograms of $\Delta z/(1+z_{\rm spec})$ in various intervals of $z_{\rm 850}$ magnitude.

\begin{figure*}[th]
\center
\includegraphics[angle=0,scale=.60]{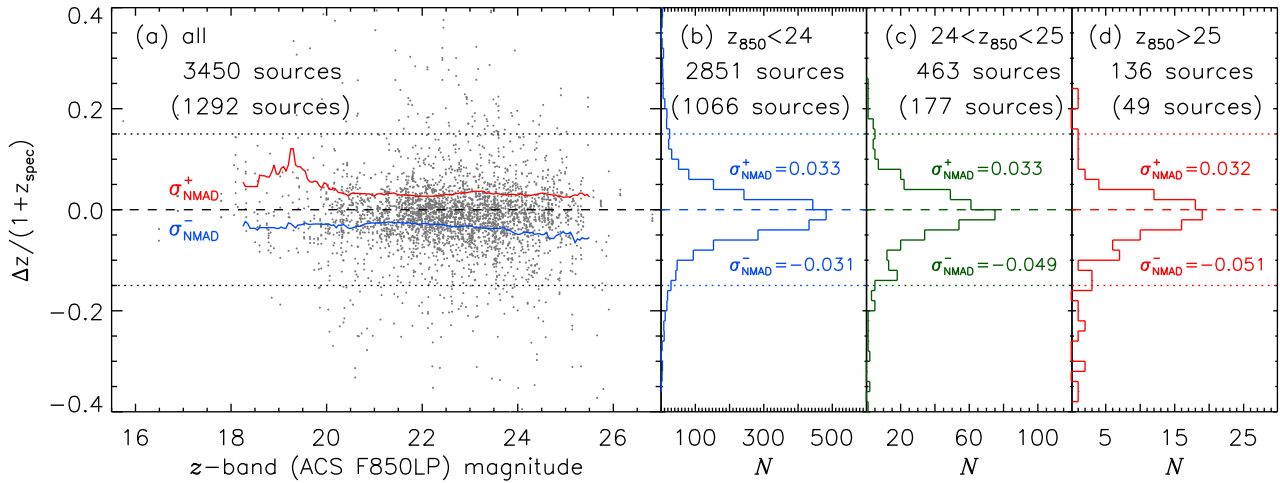}
\caption{Same as Figure~\ref{fig:zphot}, but derived with the utilization of faked photometry (corresponding to case~iv; see \S~\ref{sec:z}).} 
\label{fig:zphot2}
\end{figure*}

Strictly speaking, the above blind-test analysis of $z_{\rm phot}$ 
quality is really only applicable for $z_{\rm 850}\lsim 25.2$~mag
(the rough limit of the spectroscopic data available),
given that some of the bandpasses used in the analysis 
have dropped toward fainter magnitudes.\footnote{We examined the mean
number of detection bands ($N_{\rm filter}$) as a function of $z_{\rm 850}$
magnitude for all the 32,508 sources in the Dahlen et~al. (2010) catalog
(cf. Fig.~8b of Luo et~al. 2010).
We find that $N_{\rm filter}$ is no less than $\approx10$ 
for sources with $z_{\rm 850}<25$~mag,
while $N_{\rm filter}$ drops from $\approx10$ to $\approx7$
as $z_{\rm 850}$ goes from $\approx25$~mag to $\approx28$~mag.}
To explore effectively the true behavior of $z_{\rm phot}$ quality
at fainter magnitudes, we performed 
four additional series of blind tests (denoted as \hbox{cases i--iv})
that are almost identical to the previous blind tests with the only 
difference being the utilization of ``faked'' photometry
in four different ways.
We faked the photometry of the $z_{\rm spec}$ subsample as follows.
For each $z_{\rm spec}$ source, we first randomly picked 
a faint (i.e., $z_{\rm 850}>25$~mag), non-$z_{\rm spec}$ source
either from the Dahlen et~al. (2010) catalog (a total of 32,508 sources)
or from Sample~D (a total of 6845 sources,
whose stacked \hbox{6--8 keV} emission can account entirely for the 
unresolved \hbox{$\approx 20$--25\%} of the \hbox{6--8 keV} XRB;
see \S~\ref{sec:results} for details). 
We then applied the band coverage of the randomly picked
$z_{\rm phot}$ source to the $z_{\rm spec}$ source.
Specifically, for each filter considered,
(1) if the $z_{\rm phot}$ source was not observed, we then set the
$z_{\rm spec}$ source as non-observed; 
(2) if the $z_{\rm phot}$ source was not detected 
(i.e., upper limits applied), 
we then either set the $z_{\rm spec}$ source as non-observed 
(this corresponds to the worst scenario where all information was discarded)
or set the $z_{\rm spec}$ source as non-detected (i.e., 
we added a random $1\sigma$ fluctuation to the flux of 
the $z_{\rm spec}$ source to simulate the photometric quality of
the $z_{\rm phot}$ source and treated the derived flux 
as an upper limit); and
(3) if the $z_{\rm phot}$ source was detected, we then
did nothing with the photometry of the $z_{\rm spec}$ source.
The various combinations of parent sample (the Dahlen et~al. catalog
vs. Sample~D) and photometry treatment of the $z_{\rm spec}$ source
(non-observed vs. non-detected, when the $z_{\rm phot}$ source was 
not detected) lead to \hbox{cases i--iv}.
Table~\ref{tab:blind} shows the $z_{\rm phot}$ quality blind-test results
for \hbox{cases i--iv}
where ``faked'' photometry was utilized.
As an example, Figure~\ref{fig:zphot2} shows the 
results for case~iv.
Overall, the obtained $\sigma_{\rm NMAD}$ ranges from
0.049 to 0.055 (cf. $\sigma_{\rm NMAD}=0.043$ obtained in 
previous blind tests) and the  
outlier fraction ranges from 10.4\% to 13.1\% 
(cf. an outlier fraction of 7.1\% obtained previously).
These analyses suggest that in our case
the reduction of bandpass coverage at
$z_{\rm 850}>25$~mag does degrade the $z_{\rm phot}$ quality
to some degree, which is expected, but not severely overall.

\begin{table*}
\center
\caption{$z_{\rm phot}$ Blind-Test Results with the Utilization of Faked Photometry}
\begin{tabular}{ccccc}\hline\hline
Case & Parent Sample & Photometry Treatment & $\sigma_{\rm NMAD}$ & Outlier Fraction \\ \hline
i   & Dahlen et al. catalog & Non-observed & 0.055 & 13.1\% \\
ii  & Dahlen et al. catalog & Non-detected & 0.053 & 11.5\% \\
iii & Sample D & Non-observed & 0.051 & 11.3\% \\
iv  & Sample D & Non-detected & 0.049 & 10.4\% \\\hline
\end{tabular}
\label{tab:blind}
\end{table*}

We also made different versions of Figs.~\ref{fig:zphot} 
and~\ref{fig:zphot2} using only $z_{\rm spec}>1$ sources.
We find that these versions resemble the original Figs.~\ref{fig:zphot} 
and~\ref{fig:zphot2} closely in terms of values of $\sigma^+_{\rm NMAD}$/$\sigma^-_{\rm NMAD}$
and outlier fractions.
%
This analysis shows that there is no apparent degradation of
our $z_{\rm phot}$ quality toward high redshifts.

We then compared our $z_{\rm phot}$'s with other photometric-redshift
catalogs in this and other fields.
In general, our $z_{\rm phot}$ quality (in terms of $\sigma_{\rm NMAD}$
and outlier fraction) is consistent with that of Cardamone
et~al.~(2010), Dahlen et~al.~(2010), Luo et~al.~(2010), Rafferty
et~al.~(2011), and Salvato et~al. (2011) at similar magnitudes
(the first four catalogs have \cdfs\ coverage while the fifth one is in the COSMOS field).
Further source-to-source comparison with any of the four \cdfs\ catalogs
reveals no strong systematic difference in the $z_{\rm phot}$'s at any
magnitudes, and the typical difference in the $z_{\rm phot}$ estimates
is at the same level as the reported error bars [$\mbox{median}(\left|
z_{\rm phot,Xue}-z_{\rm phot,other} \right|) \sim \sigma_{\rm zphot} \sim 0.15$].

Overall 779 out of the 18,035 sources (4.3\%) in Sample~A
have secure $z_{\rm spec}$'s, while
Sample~D contains 537 sources (7.8\%) with a secure $z_{\rm spec}$ out of its total 6845 sources.
For Sample~D sources, 
we find acceptable agreement between our $z_{\rm phot}$'s 
and those of Dahlen et~al.~(2010) ($\sigma_{\rm NMAD}=0.080$), 
despite the different methodologies adopted 
and the challenging nature of deriving $z_{\rm phot}$'s for these 
faint sources (see \S~\ref{sec:results}).

\subsection{Rest-Frame Absolute Magnitudes}\label{sec:mags}

We followed the procedure in \S~3.1 of X10 to derive rest-frame absolute magnitudes
for each Sample~A source up to its reddest rest-frame detection band  
($K$-, $H$-, $J$-, $I$-, $R$-, $V$-, or \hbox{$B$-band}).
Briefly, we adopted the approach of template SED fitting,
which has the advantage of potentially reducing catastrophic failures
in cases of limited/incomplete photometric coverage, as opposed to
the approach of linear or log-linear interpolation/extrapolation based on photometric data.
With the input of the aforementioned Dahlen et~al. (2010; 12-band photometry) and CANDELS
($Y$, $J$, and $H$) photometry as well as
the improved set of templates (see \S~\ref{sec:z}),
we utilized ZEBRA to identify the best-fit template for each source
by fixing the source redshift to the corresponding $z_{\rm spec}$ (if available) or $z_{\rm phot}$.
We then derived rest-frame absolute magnitudes for each source based on the best-fit template.
Dust extinction is folded into our galaxy SED templates 
(see \S~\ref{sec:z}), so the derived
rest-frame absolute magnitudes are not extinction corrected.

\subsection{Stellar Masses}\label{sec:mass}

We adopted the approach described in \S~3.2 of X10 to derive stellar masses ($M_\star$) 
for the Sample~A sources.
Using the tight correlations between rest-frame optical/near-infrared 
colors and stellar mass-to-light ratios
obtained by Zibetti et~al. (2009),
\begin{equation}
{\rm log}(M_{\lambda,\star}/M_\odot)={\rm log}(L_\lambda/L_{\lambda,\odot})+b_\lambda(M_{\rm B}-M_{\rm V})+a_\lambda+0.20,\label{eq:ml}
\end{equation}
we estimated a set of stellar masses at various rest-frame bands (denoted as $\lambda$)
for each source (the values of the coefficients, $a_\lambda$ and $b_\lambda$,
can be found in Table~B1 of Zibetti et~al. 2009).
The above equation was derived by constructing spatially resolved maps of 
stellar-mass surface density in galaxies, based on the high-quality optical and near-infrared imaging
data of a sample of nine nearby galaxies that span a broad range of 
morphologies and physical properties (Zibetti et~al. 2009).
When deriving Eq.~\ref{eq:ml}, Zibetti et~al. (2009)
took into account the effects of dust in their models.
Thus, $L_\lambda$ and $M_{\rm B}-M_{\rm V}$ in Eq.~\ref{eq:ml} 
are the observed (dust-extincted) luminosity and rest-frame color.
A caveat pointed out by Zibetti et~al. (2009) is that
stellar masses of dusty starburst galaxies estimated using unresolved photometry
are likely underestimated by up to 40\% because
dusty regions are under-represented in the measured fluxes.
In Eq.~\ref{eq:ml} we have adjusted the normalization by 0.20~dex to account
for our adopted Salpeter (1955) IMF for 
stellar-mass estimates.\footnote{In this paper, we have adopted 
a conversion factor of \hbox{$\approx0.20$~dex} (i.e., \hbox{$\approx 1.6$}) between stellar masses 
estimated using a Salpeter IMF and a Chabrier IMF (with the former
stellar masses being systematically larger).}
We selected the stellar-mass estimate that corresponds to
the actual reddest rest-frame detection band of the source\footnote{Of the Sample~A (Sample~D) sources,
94.7\%, 79.5\%, and 57.0\% (94.2\%, 83.2\%, and 59.8\%)
have rest-frame $R$-band, $J$-band, and $K$-band detections (detection indicates a $>1\sigma$ signal;
see Footnote~\ref{ft:detection}) or beyond, respectively.}
because longer-wavelength (e.g., \hbox{$K$-band}) galaxy luminosities are 
much less sensitive to dust
and stellar-population effects than shorter-wavelength luminosities
(e.g., Bell \& de Jong 2000).

Using simulations, we assessed the uncertainties associated with stellar-mass estimates
that arise from our procedure for deriving photometric redshifts and
rest-frame absolute magnitudes based on template SED fitting.
For each Sample~A source, which has a photometric redshift $z_{\rm phot,i}$ and
an associated \hbox{1-sigma} error $\sigma_{\rm phot,i}$, 
we randomly drew a value $z_{\rm sim,i}$ (i.e., the simulated photometric redshift) 
from the range of $z_{\rm phot,i}\pm\sigma_{\rm phot,i}$
conservatively assuming a uniform distribution.
We then derived the simulated rest-frame absolute magnitudes for the source using
$z_{\rm sim,i}$, the aforementioned photometry, and the improved set of templates
(see \S~\ref{sec:z}), following the procedure detailed in \S~\ref{sec:mags}.
Finally, we obtained a simulated stellar-mass estimate for the source
using Eq.~\ref{eq:ml}.
For Sample~A sources, we found no systematic offset between the set of
simulated stellar-mass estimates ($M_{\star,\rm sim}$) and the set of
real stellar-mass estimates ($M_\star$), i.e., the distribution of the
logarithmic ratio between these two sets of stellar-mass estimates
[$R=\log(M_{\star,\rm sim}/M_\star$)] is symmetric and peaks at zero;
furthermore, the scatter of $R$ is 0.22~dex and largely independent of
stellar mass.
Given that the photometric-redshift errors derived with ZEBRA
generally underestimate the true errors by a factor of \hbox{$\approx
  3$--6} (see, e.g., \S~3.4 of Luo et~al. 2010), we repeated the above
simulation four times by randomly drawing $z_{\rm sim,i}$ from the
range of $z_{\rm phot,i}\pm n\sigma_{\rm phot,i}$ (where $n=3$, 4, 5,
and 6) assuming a uniform distribution.
In these four additional simulations, we also found no systematic
offset between $M_{\star,\rm sim}$ and $M_\star$ values; the scatter
of $R$ is 0.40, 0.45, 0.50, and 0.56~dex for $n=3$, 4, 5, and 6,
respectively.
We expect to have smaller scatters in $R$ if we randomly
draw $z_{\rm sim,i}$ assuming a Gaussian distribution that peaks at
$z_{\rm phot,i}$, which is likely closer to reality.
The above analyses show that the stellar-mass errors produced 
by the uncertainties of photometric redshifts and rest-frame absolute magnitudes
are typically smaller than $\approx 0.2$--$0.5$ dex.

\begin{figure}[t]
\center
\includegraphics[angle=0,scale=.35]{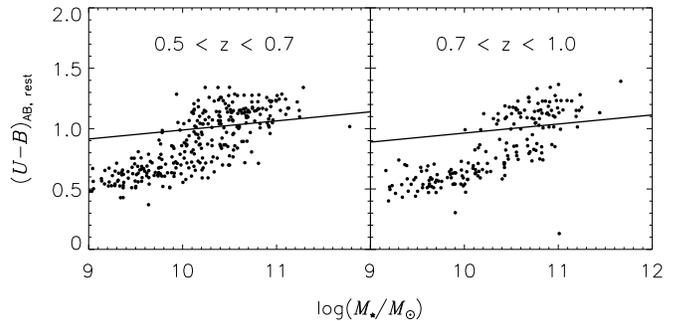}
\caption{Color-mass distribution plots, for direct comparison with the
  two bottom panels in Fig.~4 of Peng et~al.~(2010) to evaluate the validity 
  of our mass estimates. The $y$-axis is rest-frame $U-B$ color
  (converted into the AB magnitude system), and the $x$-axis is the
  logarithm of our mass estimate reduced by $0.2$~dex to adjust for the
  offset between our IMF (Salpeter) and the Chabrier IMF adopted in
  Peng et~al.~(2010). The solid line is the division between red and
  blue galaxies used by Peng et~al.~(2010), which is a function of
  $U-B$, mass, and redshift (see their Eq.~2).  The sources used in
  these two panels are in the same redshift intervals as used in the
  two Peng et~al.~(2010) panels, and we also applied a magnitude cut
  of $i<22.5$ since their catalog is flux-limited at $I<22.5$.}
\label{fig:colormass}
\end{figure}

We assessed the robustness of our stellar-mass estimates through
several checks.
First, we compared our stellar-mass estimates with those presented by
X10 and Mullaney et~al. (2012).  For all 
sources in X10 and Mullaney et al.~(2012), we find general agreement between common
sources, with a median ratio of $\approx 1.0$ between the two
estimates (after taking into account different choices of IMFs and
rest-frame bands that are used for stellar-mass estimates) and $\lsim
0.35$~dex random scatter.
Second, we compared our galaxy stellar-mass distributions to those in the COSMOS field.
Ilbert et~al. (2010) computed the stellar masses of the COSMOS galaxies
where sources with $i^+<25.0$ have the most reliable photometric redshifts and mass estimates.
For each of the chosen Subaru $i^+$-band limiting magnitudes
(i.e., $i^+_{\rm limit}=22.5$, 23.0, 23.5, 24.0, 24.5, and 25.0),\footnote{For each Sample~A source, we utilized a
  $K$-correction package (kcorrect.v4\_1\_4; Blanton \& Roweis 2007)
  to convert the $z_{\rm 850}$ magnitude into the $i^+$
  magnitude by convolving the best-fit SED template of the source with
  the $z_{\rm 850}$ and $i^+$ filter curves and computing the
  differences between the derived $z_{\rm 850}$ and $i^+$
  magnitudes (typically $|z_{\rm 850}-i^+|<0.5$~mag).}             
the distributions of our stellar masses and the Ilbert et~al.~(2010) stellar masses
are generally similar, with comparable median stellar masses and Kolmogorov-Smirnov
test probabilities ranging from 8.0\% to 49.4\% 
that indicate similar stellar-mass distributions. 
Third, we compared our color-mass distribution with that in Peng
et~al.~(2010). Figure~\ref{fig:colormass} is our color-mass diagram,
which is in parallel with the two bottom panels in Fig.~4 of Peng
et~al.~(2010). The color-mass bimodality feature and the distribution of
the sources in the color-mass plane in our Fig.~\ref{fig:colormass}
are very similar to those in Fig.~4 of Peng et~al.~(2010).
Finally, we also estimated stellar masses utilizing 
the Fitting and Assessment of Synthetic Templates (FAST;
Kriek et~al.~2009) package that is based on galaxy SED fitting.
We adopted the same stellar population synthesis model, IMF,
dust-extinction law, star-formation history, and metallicity ranges
as those described in \S~\ref{sec:z} to ensure consistency between
estimates of $z_{\rm phot}$, rest-frame absolute magnitudes, and stellar masses. 
We found that stellar masses calculated by FAST
are consistent with those based on Zibetti et~al.~(2009) 
after taking into account different choices of IMFs, with an RMS of
\hbox{$\lsim 0.4$ dex}, which is the typical precision of such methods.
Throughout this paper, we have chosen to adopt stellar masses based on Zibetti
et~al.~(2009) that are more directly related to source colors and
rest-frame absolute magnitudes and thus less dependent on model and parameter choices.
We have verified that the same basic results presented below can be obtained by using stellar masses
calculated with FAST.

\section{Analysis and Results}\label{sec:results}

In Table~\ref{tab:data} we present the derived source properties of the 18,035 Sample~A sources.
As discussed in \S~\ref{sec:intro}, mounting evidence has shown that
luminous AGNs tend to reside in massive (i.e., $M_\star\gsim 10^{10}\:M_\odot$) and red galaxies over at least the last $\approx 80$\% 
of cosmic history, i.e., \hbox{$z\approx 0$--4}
(e.g., Barger et~al. 2003; Bundy et~al. 2008; Brusa et~al. 2009; Silverman et~al. 2009; X10; Mullaney et~al. 2012).
Therefore, we utilize these mass and color constraints as clues to search for
the underlying population of luminous but highly obscured AGNs that
are responsible for the unresolved \hbox{$\approx 20$--25\%} of the \hbox{6--8 keV} XRB.

\begin{table*}
\caption{Derived Properties for the Sources in Sample A}
\resizebox{\textwidth}{!}{%
\begin{tabular}{rcccccccccccr}\hline\hline
No. & RA & DEC & $z_{\rm spec}$ & $z_{\rm phot}$ & $z_{\rm ph,low}$ & $z_{\rm ph,up}$ & $M_{\rm U}$ & $M_{\rm B}$ & $M_{\rm V}$ & $M_{\rm reddest}$ & F$_{\rm reddest}$ & log($M_\star/M_\odot$) \\
(1) & (2) & (3) & (4) & (5) & (6) & (7) & (8) & (9) & (10) & (11) & (12) & (13)\\\hline
... ...\\
 9000 & 53.11829 & $-$27.86707 & $-$1.000 & 1.620 & 1.548 & 1.707 & $-$17.88 & $-$17.98 & $-$18.42 & $-$20.15 & 6 &  8.88 \\
 9001 & 53.11829 & $-$27.84127 & $-$1.000 & 0.610 & 0.594 & 0.670 & $-$15.07 & $-$14.97 & $-$15.59 & $-$19.27 & 7 &  8.60 \\
 9002 & 53.11832 & $-$27.81530 & $-$1.000 & 0.550 & 0.533 & 0.665 & $-$15.67 & $-$14.98 & $-$15.21 & $-$17.07 & 7 &  7.26 \\
 9003 & 53.11834 & $-$27.72375 & $-$1.000 & 0.395 & 0.362 & 0.613 & $-$14.85 & $-$15.07 & $-$15.63 & $-$17.97 & 7 &  8.02 \\
 9004 & 53.11835 & $-$27.70741 & $-$1.000 & 0.750 & 0.605 & 0.863 & $-$15.94 & $-$15.77 & $-$15.90 & $-$16.42 & 4 &  7.60 \\
... ...\\
\hline
\end{tabular}}
\\Notes. The full table contains 18,035 entries and 20 columns for each entry.
Columns:
(1) Source sequence number (from 1 to 18035).
(2, 3) J2000 right ascension and declination (in degrees).
(4) Spectroscopic redshift ($-1.000$ indicates no spectroscopic redshift available).
(5) Photometric redshift.
(6, 7) 1-$\sigma$ lower and upper limits on photometric redshift.
(8, 9, 10) Rest-frame absolute $U$-, $B$-, and $V$-band magnitude (Vega mags).
(11) Rest-frame absolute magnitude (Vega mags) that corresponds to the reddest rest-frame coverage.
The conversion between Vega and AB rest-frame absolute magnitudes is: 
$M_{\rm AB}=M_{\rm Vega}+m_{\rm conv}$, where $m_{\rm conv}=
0.628/-0.102/0.029/0.264/0.501/0.914/1.381/1.839$ for 
rest-frame $U/B/V/R/I/J/H/K$-band, respectively;
we derived these $m_{\rm conv}$ values using a $K$-correction package
(kcorrect.v4\_1\_4; Blanton \& Roweis 2007).
(12) Flag of reddest rest-frame coverage. This flag shows which stellar-mass estimate is adopted as the final estimate (see \S~\ref{sec:mass}). F$_{\rm reddest}=$(1, 2, 3, 4, 5, 6, 7) means that
$M_{\rm reddest}=M_{\rm B/V/R/I/J/H/K}$ and $M_{\rm B/V/R/I/J/H/K,\star}$ is adopted, respectively.
(13) Stellar-mass estimate adopted in this paper.
(14) Stellar-mass calculated by FAST.
(15, 16) CANDELS $Y$-band magnitude and associated 1-$\sigma$ uncertainty (AB mags; $-99.00$ indicates no photometry available for this filter).
(17, 18) CANDELS $J$-band magnitude and associated 1-$\sigma$ uncertainty (AB mags).
(19, 20) CANDELS $H$-band magnitude and associated 1-$\sigma$ uncertainty (AB mags).\\
(This table is available in its entirety in a machine-readable form in the online journal. A portion is shown here for guidance regarding its form and content.)
\label{tab:data}
\end{table*}

\begin{figure*}[t]
\center
\includegraphics[angle=0,scale=.8]{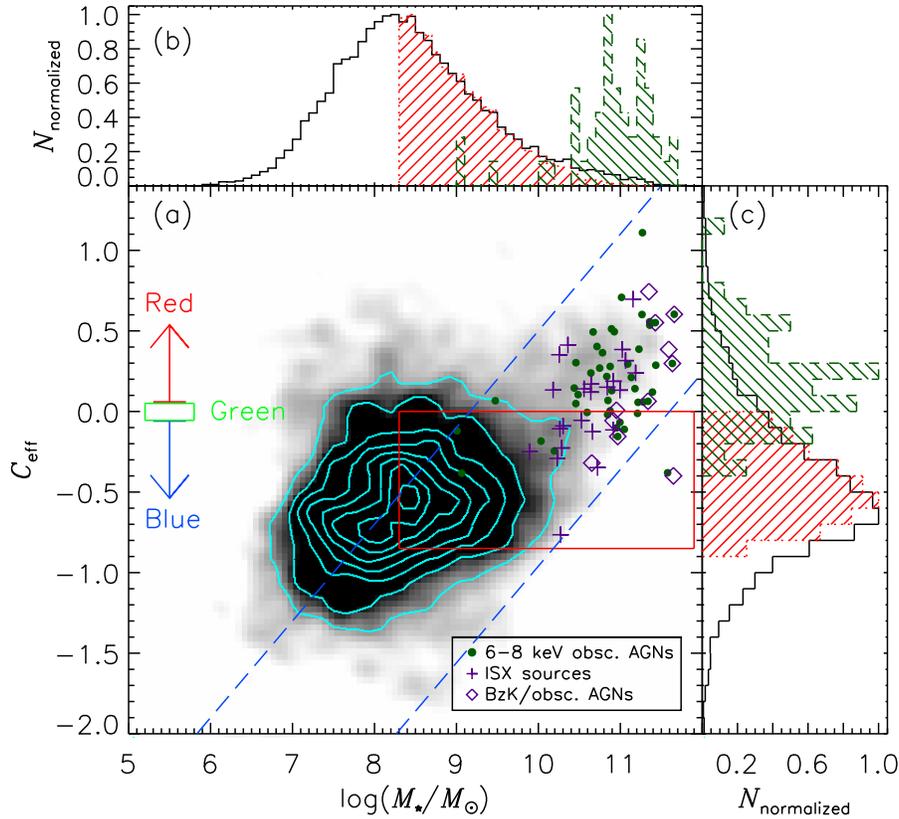}
\caption{(a) Effective color-mass diagram for Sample~A, which is shown as a density map overlaid with
contours (the 8 contour levels are 3, 6, 9, 12, 15, 18, 21, and 24 sources per pixel).
The large red rectangle highlights the region occupied by the Sample~D sources.
The two parallel, diagonal, long-dashed lines enclose 
a subsample of sources (discussed in \S~\ref{sec:robustness}) 
that lie within a diagonal stripe.
For comparison, 
a sample of 47 highly obscured AGNs detected in the \hbox{6--8~keV} band in the 4~Ms \cdfs\ (large dark green filled circles; X11),
a sample of 23 \xray\ undetected and infrared-selected highly obscured
AGN candidates (crosses; Luo et~al. 2011), and
a sample of 11 highly obscured AGNs that were $K<22$ $BzK$-selected
galaxies (diamonds; Alexander et~al. 2011)
are also plotted (see \S~\ref{sec:results}).
The division scheme of the red sequence, the green valley, and the blue cloud is 
illustrated on the left side.
(b) Normalized (peaking at unity) stellar-mass histograms for Sample~A
(black histogram), Sample~D (red shaded histogram), and the
sample of 47 highly obscured, \hbox{6--8~keV} detected
AGNs (dark green shaded histogram; for clarity, we do not show the histograms for the aforementioned 
23 highly obscured AGN candidates and 11 highly obscured AGNs).
(c) Same as Panel (b), but for normalized effective-color histograms.
}
\label{fig:cmd}
\end{figure*}

Figure~\ref{fig:cmd}a shows the effective color vs. mass diagram for Sample~A 
(shown as a density map overlaid with contours),
where the effective color is defined as
\begin{equation}
C_{\rm eff}=(U-V)_{\rm rest}+0.31z+0.08M_{\rm V}+0.51.
\label{eq:ceff}
\end{equation}
In Eq.~\ref{eq:ceff}, $(U-V)_{\rm rest}$ is the rest-frame $U-V$ color (i.e., $M_{\rm U}-M_{\rm V}$),
$z$ is the redshift, and $M_{\rm V}$ ($M_{\rm U}$) is 
the rest-frame absolute $V$-band ($U$-band) magnitude.
The definition of $C_{\rm eff}$ is based on the equation separating galaxies into
the red sequence and the blue cloud of Bell et~al. (2004),
who studied the color distribution of \hbox{$\approx 25$,000} $R\lsim 24$ galaxies
with $0.2<z\le 1.1$.
Taking into account a typical color scatter of $\lsim0.2$~mag for the red-sequence 
color-magnitude relation (see \S~4 of Bell et~al. 2004),
the Bell et~al. equation separates blue and red galaxies reasonably well down to fainter 
magnitudes out to \hbox{$z\approx 3$--4} (e.g., X10), and thus we 
use $C_{\rm eff}$ to define whether a galaxy in Sample~A is red or not\footnote{Galaxies
in the red sequence, the green valley, and the blue cloud have $C_{\rm eff}\geq 0.05$,
$-0.05<C_{\rm eff}<0.05$, and $C_{\rm eff}\leq -0.05$, respectively,
given that 0.05 is the typical
``half-width'' of the green valley in a color-magnitude diagram
(e.g., Nandra et~al. 2007; X10).
By definition, we would expect the $C_{\rm eff}$ distribution to be double-peaked
(i.e., red and blue peaks), which is, however, not clearly seen in Fig.~\ref{fig:cmd}c
(black histogram) due to the dilution caused by
color errors,
uncertainties in redshift estimates, 
and large numbers of low-mass blue galaxies.
Indeed, if we consider only, e.g., galaxies with $0<z<1$ and $M_\star>10^{9.5}\:M_\odot$ (as in X10),
then color bimodality is clearly seen.}
given that the sources in Sample~A span a wide range in redshift and luminosity. 
As shown in Fig.~\ref{fig:cmd}a, there is a correlation between
stellar mass and effective color (with large scatter) such that more massive
galaxies are generally redder (i.e., having larger $C_{\rm eff}$ values),
consistent with previous results.
The normalized histograms of stellar mass and effective color are shown in
Figs.~\ref{fig:cmd}b and \ref{fig:cmd}c (black histograms),
respectively.
For comparison, three additional samples of highly obscured AGNs or
AGN candidates are also plotted on Fig.~\ref{fig:cmd}a.
The sample labeled with dark-green filled circles consists of 47~highly
obscured AGNs at \hbox{$z\approx 0.5$--3} detected in the \hbox{6--8~keV} band in the central $6\arcmin$~area of the 4~Ms
\cdfs\footnote{These 47~AGNs are the 4~Ms \cdfs\ main-catalog sources
that have an effective photon index of $\Gamma_{\rm eff}\le 1.0$ and 
satisfy a binomial-probability source-selection criterion of $P<0.004$ in the \hbox{6--8~keV} band.
In X11, the $P<0.004$ source-detection criterion was applied only in the
\hbox{0.5--8}, \hbox{0.5--2}, and \hbox{2--8~keV} bands; 
here we extended the use of this criterion for \hbox{6--8~keV} source detection.} 
(their mass and color histograms are shaded in dark green in
Figs.~\ref{fig:cmd}b and \ref{fig:cmd}c; X11).
The points labeled with crosses are a sample of 
23~highly obscured AGN candidates at \hbox{$z\approx 0.5$--1}
that were X-ray undetected and selected by their infrared
star-formation rate (SFR) excess (i.e., infrared-based SFRs being a factor of $\geq$3.2
higher than SFRs determined from the UV after correcting for dust
extinction; Luo et~al. 2011).
The points labeled with diamonds are a
sample of 11~AGNs at \hbox{$z\approx 2$} that were $K<22$ $BzK$-selected
galaxies and identified as highly obscured using the 4~Ms
\cdfs\ data (Alexander et~al. 2011).\footnote{The majority of the 
highly obscured AGNs mentioned here have $L_{\rm 0.5-8\;keV}<10^{43.7}$~\lum, which indicates that
their hosts dominate the optical-to-near infrared emission thus ensuring 
reliable estimates of host stellar masses and colors (see \S~4.6.3 of X10 for details).}
As expected, the vast majority of these sources are massive and on the red sequence,
the green valley, or the top of the blue cloud.

We then proceeded to stack different sub-groups of Sample A to
investigate which sources produce the majority contribution to the
unresolved \hbox{$\approx 20$--25\%} of the \hbox{6--8~keV} XRB.
We adopted the same stacking procedure as detailed in \S~3.1 of Luo
et~al. (2011). Briefly, total counts (including background) for each
individual source were extracted from an aperture 3\arcsec\ in diameter
centered on its optical position.
Background counts for each source were estimated by taking the mean of
the counts within 1000 apertures (also with 3\arcsec\ diameter each), which
were randomly placed within a 1\arcmin-radius circle around the source
avoiding any known X-ray source (i.e., outside of twice the \hbox{0.5--2 keV} 90\%
encircled-energy aperture radius of any 4~Ms \cdfs\ main-catalog source).
Stacked counts [total ($S$) or background ($B$)]
were the summation of counts
from the stacked sample with proper aperture correction applied.
The net source counts are then given by $S-B$, and the S/N is calculated as
$(S-B)/\sqrt{B}$ where Gaussian statistics are assumed given 
the large values of $S$ and $B$.

\begin{figure}[t]
\includegraphics[angle=0,scale=.47]{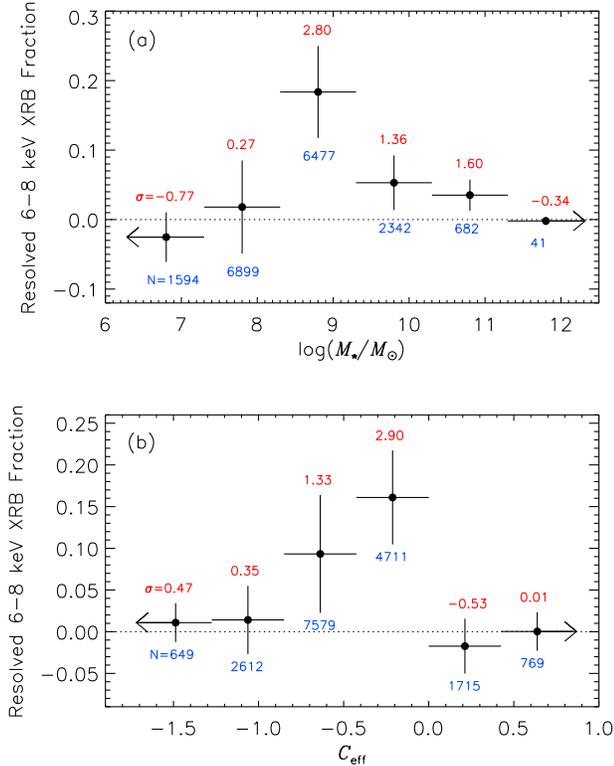}
\caption{(a) Resolved 6--8~keV XRB fractions for Sample~A sources in various stellar-mass bins.
The number of sources ($N$) and the significance (in terms of $\sigma$) of the stacked signal
in each stellar-mass bin are annotated accordingly.
The horizontal dotted line indicates zero resolved 6--8~keV XRB fraction.
(b) Same as panel (a), but for effective-color bins.
Here the quoted significances are in general low due to
the dilution of signal caused by sample splitting into many stacking bins
(this also applies to Fig.~\ref{fig:prop}).
}
\label{fig:cases}
\end{figure}

Motivated by the fact that most of the obscured AGNs (and AGN candidates) appear in
the massive and relatively red corner of Fig.~\ref{fig:cmd}a, we divided Sample A
into various stellar-mass and effective-color bins and stacked the
sources in each bin. 
Figures.~\ref{fig:cases}a and \ref{fig:cases}b show the stacking results,
and Table~\ref{tab:samples} gives some detailed stacking results
(e.g., stacked net counts, stacked signal-to-noise ratio,
effective photon index, and resolved \hbox{6--8 keV} XRB fraction) 
for some cases of interest.
It seems clear that
the \hbox{$\approx 20$--25\%} unresolved \hbox{6--8 keV}
XRB mostly lies in galaxies with $M_\star\ge 2\times 10^8\:M_\odot$
(i.e., Sample~B; see Table~\ref{tab:samples}), in particular in the
bin of $2\times 10^8\le M_\star/M_\odot\le 2\times 10^9$ where a
$2.8\sigma$ signal was obtained.
Moreover, the signal also mostly arises from the Sample~A sources on the top of the blue cloud,
i.e., the unresolved \hbox{6--8 keV} XRB has major contributions
from galaxies with $-0.85<C_{\rm eff}<0$ (i.e., Sample~C; see Table~\ref{tab:samples}),
in particular in the
bin of $-0.45<C_{\rm eff}<0$ where a
$2.9\sigma$ signal was obtained.

By applying both the mass and color constraints together (i.e., selecting the common sources
in Sample~B and Sample~C),
we obtained Sample~D (i.e., $M_\star\ge 2\times 10^8\:M_\odot$ and $-0.85<C_{\rm eff}<0$;
see Table~\ref{tab:samples}).
There are 6845 galaxies in Sample~D that can account entirely for 
the unresolved \hbox{$\approx 20$--25\%} of the
\hbox{6--8~keV} XRB (see Table~\ref{tab:samples}), and the stacked 
\hbox{6--8~keV} signal from these galaxies
is significant at the $3.9\sigma$ level (corresponding to a chance of $p=5.1\times 10^{-5}$
that the signal was generated by Poisson noise).
The region occupied by the Sample~D sources in the effective color-mass diagram
is highlighted with a large red rectangle in Fig.~\ref{fig:cmd}a,
and their normalized stellar-mass and effective-color distributions are shown in
Figs.~\ref{fig:cmd}b and \ref{fig:cmd}c, respectively.
Additional information about the stacking results for various samples
can be found in Table~\ref{tab:resolved} and Fig.~\ref{fig:xrbfrac}.

\begin{figure}[th]
\includegraphics[angle=0,scale=0.47]{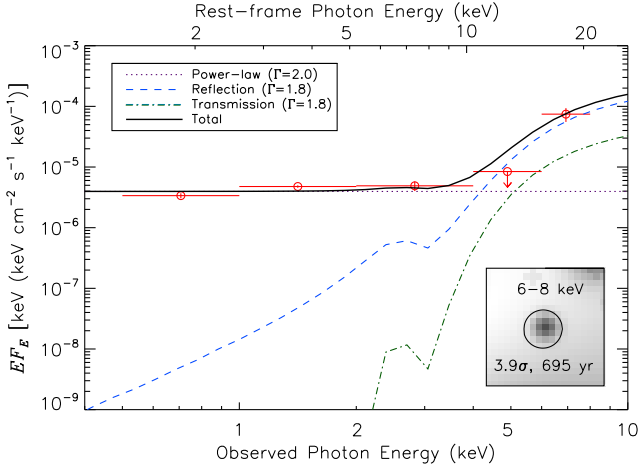}
\caption{Stacked \xray\ spectrum (open circles) for the 6845 sources in Sample~D
(the top $x$-axis shows the rest-frame photon energy at $z=1.6$, which is the
median redshift of the Sample~D sources; see Table~\ref{tab:samples}). 
The downward arrow in the \hbox{4--6~keV} band indicates a $3\sigma$ upper limit.
The solid curve is a schematic fit to the stacked \xray\ spectrum,
which is the sum of three components (each evaluated at $z=1.6$):
an unabsorbed power-law component accounting for star formation (dotted line; $\Gamma=2.0$),
a pure reflection component from the AGN (dashed curve),
and a pure transmission component from the AGN (dashed-dot curve).
{\it Inset}: Stacked, adaptively smoothed, \hbox{6--8~keV} image,
with the 3\arcsec\ diameter photometric aperture, the significance of the stacked signal, and the total stacked exposure shown.}
\label{fig:spectrum}
\end{figure}

Figure~\ref{fig:spectrum} shows the stacked \hbox{0.5--8~keV} spectrum for the Sample~D sources
(open circles),
with the stacked, adaptively smoothed \hbox{6--8~keV} image shown as the inset.
The facts that the \hbox{6--8~keV} flux is significantly elevated 
(compared to the fluxes in the lower-energy bands)
and that there is no detection in the \hbox{4--6~keV} band suggest that
highly obscured AGNs dominate the stacked spectrum at high \xray\ energies 
(see \S~\ref{sec:constraints} for more discussion).
The apparent inconsistency between the hard stacked spectrum 
(see Fig.~\ref{fig:spectrum}) and the relatively small band ratio
(0.48, corresponding to $\Gamma_{\rm eff}=1.60$;  
see Table~\ref{tab:samples}) for Sample~D is due to the fact that
the observed \hbox{2--8 keV} count rate, which dilutes the 
\hbox{6--8 keV} contribution, is used for the calculation. 
We stress that this observed spectral rise 
at the \hbox{6--8~keV} band is not
caused by our sample selection, based on the following analyses:
(1) The ratio between the stacked \hbox{6--8} and \hbox{4--6 keV} fluxes
for the parent sample, i.e., Sample~A, is larger than 9, meaning that
the spectral rise observed in the stacked Sample~D spectrum is
actually intrinsic to Sample~A (see Fig.~\ref{fig:xrbfrac});
(2) We stacked Sample~D and non-Sample~D sources in the \hbox{4--6 keV} band
respectively and do not find any difference between the average
\hbox{4--6 keV} fluxes for these two samples;
(3) We examined the resolved \hbox{4--6 keV} XRB fractions
for Sample~A sources in various stellar-mass and effective-color bins
(cf. Fig.~\ref{fig:cases}) and do not find any
correlation or pattern between the resolved \hbox{4--6 keV} XRB fractions
and stellar masses/effective colors 
(unlike the case of Fig.~\ref{fig:cases}),
which suggests that, by applying our stellar-mass and effective-color
cuts, we did not discard sources that have
a relatively larger \hbox{4--6 keV} flux.

\section{Robustness of Stacking Results}\label{sec:robustness}

It is important to assess the robustness of our stacking results and
the significance of our stacked signal.
Our stacking strategy, i.e., selecting sources in $M_\star$ and
$C_{\rm eff}$ space, is strictly physically motivated, although the
exact threshold values of $M_\star$ and $C_{\rm eff}$ were chosen for
a yield of strong signal. As a result, it is possible that the
significance value of $3.9\sigma$ reported above is somewhat
overestimated.
Therefore, we performed 1000 10-fold cross-validation tests 
(Efron \& Tibshirani 1993; Kohavi 1995; Davison \& Hinkley 1997) to assess
further the significance of our stacked signal. In each of the 
10-fold cross-validation tests, we randomly split the data (i.e., Sample~A) 
into ten subsamples of the same size.
Taking one of the subsamples as the testing data, we used the rest of the
subsamples as the training data upon which exploratory sample-selection
criteria were utilized to find the best threshold values for $M_\star$ and
$C_{\rm eff}$ in a similar way to the construction of Sample~D.\footnote{We
required that the grid in $\log{M_\star}$-$C_{\rm eff}$ space used to 
find the threshold values be not finer than $0.3$~dex for
$\log{M_{\star}}$ or $0.05$ for $C_{\rm eff}$,
which are the typical errors on $M_\star$ or $C_{\rm eff}$.}
These threshold values were then used for selecting sources from the
testing subsample. The process was repeated with each of the ten
subsamples being the testing data once, and in the end all the
selected sources in these ten folds were combined for stacking,
which provided one estimate of the significance of our stacked signal
($\sigma_{\rm s}$).
The 10-fold cross-validation test was performed $1000$ times, and a
distribution for $\sigma_{\rm s}$ was obtained. The median value for 
$\sigma_{\rm s}$
is $3.3$ (corresponding to a chance of $p=5.0 \times 10^{-4}$ of the
signal coming from pure Poisson noise). 
This value is likely to be slightly pessimistically biased
because the effectiveness of the search for the best threshold
$M_\star$ and $C_{\rm eff}$ values is highly dependent on the training
sample size, whereas in 10-fold cross validation tests 10\% of the data were
not used for training (Kohavi 1995). Therefore, the true significance
of our stacked signal should be between $3.3\sigma$ and $3.9\sigma$.

We also performed several other robustness and consistency tests.  For
example, we stacked randomly selected sets of 6845 sources from
Sample~A.  We repeated this procedure 10,000 times and find not a
single case where the stacked \hbox{6--8~keV} signal resolves more XRB
or has higher significance compared to Sample~D. This is consistent
with our reported significance of ($3.3$--$3.9$)$\sigma$ with
$p=(5.1$--$50) \times 10^{-5}$.
We also stacked all the non Sample~D sources (i.e., $18035-6845=11190$
sources) and find that the stacked \hbox{6--8~keV} signal is, as
expected, consistent with background (a $0.1\sigma$ signal).
We furthermore investigated the effects of changes in the
sample-selection criteria on our stacking results.  We considered
various combinations of stellar-mass and effective-color threshold
values [e.g., $M_{\star,\rm threshold}$ varying between
  \hbox{(1--5)$\times 10^8\:M_\odot$} and the lower and upper $C_{\rm
    eff}$ threshold values varying by $\pm0.2$, respectively] and
obtained similar stacking results to that of Sample~D in all cases.

Finally, we have explored another purely physically motivated stacking
strategy, which is formulated upon observation of the distribution of
the three groups of obscured AGNs or AGN candidates in
Fig.~\ref{fig:cmd}a. 
Given the locations of these obscured AGNs
or AGN candidates in the figure,
we selected sources lying within a diagonal stripe,
$5.5<0.6\log{M_\star}-C_{\rm eff}<7.0$
(i.e., the region enclosed by the two
parallel, diagonal, long-dashed lines in Fig.~\ref{fig:cmd}a). 
This stripe, running
from the upper right corner to the lower left in Fig.~\ref{fig:cmd}a,
is essentially the narrowest 
stripe containing all the obscured AGNs or AGN candidates 
except two ``outliers'' that are located around the 
$C_{\rm eff}\approx -0.4$ and $\log{M_\star}>11$ area.
There are
8415 sources within this stripe, and they contribute $25.9 \pm 7.6$\%
to the unresolved 6--8 keV XRB (a 3.5$\sigma$ stacked signal).

\begin{figure}[th]
\includegraphics[angle=0,scale=.47]{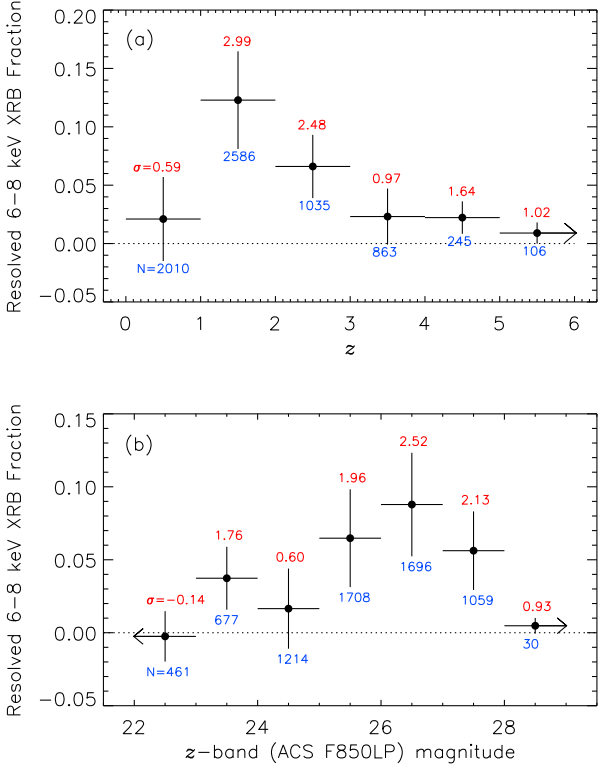}
\caption{(a) Resolved 6--8~keV XRB fractions for Sample~D sources in various redshift bins (cf.~Fig.~\ref{fig:cases}).
(b) Same as Panel (a), but for $z$-band magnitude bins.
}
\label{fig:prop}
\end{figure}

\section{Discussion}\label{sec:discussion}

\subsection{General Properties of the Galaxies Hosting the Underlying Highly Obscured AGNs}\label{sec:genprop}

Given that the Sample~D sources can account for the unresolved \hbox{$\approx 20$--25\%} of the 6--8~keV
XRB, it is of interest to determine what sources provide the majority contributions,
i.e., what sources in Sample~D are most likely to host ``hidden'' highly obscured AGNs.
We therefore examined the resolved 6--8~keV XRB fractions for the Sample~D sources in various 
redshift and $z$-band magnitude bins, as shown in Figs.~\ref{fig:prop}a and \ref{fig:prop}b,
respectively.
It appears that the galaxies with redshifts $1\lsim z\lsim 3$
(see Fig.~\ref{fig:prop}a) and magnitudes \hbox{$z_{\rm 850}\approx 25$--28} (see Fig.~\ref{fig:prop}b)
make the major contributions to the unresolved \hbox{6--8~keV}
XRB, thus being more likely to host the highly obscured AGNs that
escape from even the deepest \chandra\ observations. 
Marchesini et~al. (2012) studied the rest-frame $V$-band luminosity function of galaxies 
at $0.4\le z<4.0$.
Based on best-fit $M^*_{\rm V}$~values (for a Schechter luminosity function) in the
different redshift ranges presented in their Table~2,
we estimate that the above Sample~D sources with \hbox{$z_{\rm 850}\approx 25$--28} at $1\lsim z\lsim 3$
typically have \hbox{(0.05--$0.10)\;L^*_{\rm V}$}.
Here we do not expect cosmic variance induced 
by large-scale structures (LSS)
to affect the basic redshift distribution observed in Fig.~\ref{fig:prop}a
(thus affecting our basic results) in a significant way 
because of the following (also see \S~\ref{sec:intro} for
a brief discussion of cosmic variance):
(1) All known prominent LSS in the \cdfs\ 
has $z<1$ (i.e., $z_{\rm LSS}=0.67$ and 0.73; see, e.g.,
Silverman et~al. 2010, and their Fig.~11 where small enhancements
at other redshifts can also be seen); and
(2) The broad redshift bins ($\Delta z=1$) that we considered 
in Fig.~\ref{fig:prop}a and the broad redshift range 
($1\lsim z\lsim 3$) where we find most of the signal should be, by design,
relatively insensitive to the effects of cosmic variance induced by LSS.

As described in \S~\ref{sec:results}, three additional samples of
highly obscured AGNs or AGN candidates are also shown in the effective
color-mass diagram (Fig.~\ref{fig:cmd}a), which are seen to be massive and relatively red.
This motivates and supports our utilization
of the mass and color constraints as clues
in identifying a source population (i.e., Sample~D) responsible for
the unresolved \hbox{6--8~keV} XRB.\footnote{We note that
the \hbox{6--8 keV} signal is weak for the \xray\ undetected sources with \hbox{$M_\star>10^{10} \:M_\odot$}
(the resolved fraction is $3.0\%\pm2.8\%$ with $\sigma=1.1$), 
which could potentially be the more heavily obscured
counterparts of the \xray\ detected highly obscured AGNs (see Fig.~\ref{fig:cmd}a).
We speculate that there are simply not
enough objects to produce a significant signal for such massive galaxies.}
Nevertheless, the hosts of these highly obscured AGNs or AGN candidates are
much more massive and redder than the Sample~D sources (see
\hbox{Figs.~\ref{fig:cmd}a--\ref{fig:cmd}c}). In particular, the
stellar masses of the Sample~D sources appear notably low (most having
$2\times 10^8\lsim M_\star/M_\odot\lsim 2\times 10^9$) with a median
stellar mass of $\approx 8\times 10^8\: M_\odot$. 
However, for a typical star-forming galaxy with $M_\star=8\times 10^8\:M_\odot$
at $z=1.6$ (the median redshift of the Sample~D sources), 
its stellar mass will grow by a factor of \hbox{$\approx4$--50}
by the present day, which places its $z=0$ stellar mass at \hbox{$\approx 0.1$--1} times
the stellar mass of the Milky Way ($\approx 5\times 10^{10} \:M_\odot$; e.g., Hammer et~al. 2007).
The above estimate of mass-growth factor was made based on the calculations done by Leitner~(2012)
and equations~1 and 21 in Peng et~al. 
(2010) taking into account the effects of mergers and merger-induced quenching.
This predicted stellar-mass growth appears
consistent with the fact that Lyman-alpha emitters with a typical stellar mass of \hbox{$\sim 10^8$--$10^9 M_\odot$} at \hbox{$z\sim 2$--3} are thought to grow into galaxies about as massive as the Milky Way by the present day (Gawiser et~al.~2007; Guaita et~al.~2010).
Such a significant stellar-mass growth would imply a very large reservoir of gas present
to sustain a large amount of star formation since $z=1.6$.
In addition to supporting star formation, this gas at $z=1.6$ may also feed the
supermassive black hole (SMBH; explaining the common accretion likely present in Sample~D)
and obscure the SMBH (explaining the high obscured AGN fraction apparently seen 
in Sample~D; see \S~\ref{sec:con_frac} for details).
Most of the Sample~D sources are brighter than $z_{\rm 850}\approx27$,
thus having reasonably good photometric coverage (over 99\% of the sources in
Sample D have detections in at least 9 bands), so their
photometric-redshift and stellar-mass estimates are of sufficient quality for our study
(see \S~\ref{sec:z} and \S~\ref{sec:mass}).
The above results therefore imply that there are a significant
number of highly obscured AGNs that are hosted by relatively low-mass galaxies ($2\times
10^8\lsim M_\star/M_\odot\lsim 2\times 10^9$) at $1\lsim z\lsim 3$.\footnote{The wording of ``relatively low-mass'' here
means that the Sample~D sources have low masses
when compared to the aforementioned highly obscured AGNs or AGN candidates.
They do, of course, still have high masses when compared to the large number of non-Sample~D sources
(see Figs.~\ref{fig:cmd}a and \ref{fig:cmd}b).}

Such an AGN population might seem surprising given that the majority of the \xray\
detected AGNs reside in massive galaxies.
We thus discuss in \S~\ref{sec:constraints} constraints upon these underlying
highly obscured AGNs and their parent population.

\subsection{Constraints upon Underlying Highly Obscured AGNs and Their Parent Population}\label{sec:constraints}

\subsubsection{Spectral constraints}\label{sec:spec_cons}

At \hbox{$z=1.6$}, which is the median redshift of the Sample~D sources,
moderately Compton-thick obscuring material ($N_{\rm H}>1.5\times10^{24}$~cm$^{-2}$) would be required
to absorb \hbox{X-rays} strongly up to rest-frame $\approx 16$~keV,
but then permit higher energy emission to penetrate through the material. 
In this regime, performing an absorption correction to derive a typical
luminosity is difficult and geometry dependent (and thus subject to large uncertainties). 
However, based on results for local Seyfert galaxies (e.g., Guainazzi et~al. 2000)
and utilizing the {\sc MYTORUS} model (Murphy \& Yaqoob 2009),\footnote{We used a model that
consists of the transmitted continuum, the scattered (i.e., reflection) 
continuum, and no emission lines.
In the model we adopted
$\Gamma=1.8$, $z=1.6$, an inclination angle of 90\degree, and a varying $N_{\rm H}$
(other parameters were fixed to their default values).
We found that a column density of $N_{\rm H}\approx 4\times10^{24}$~cm$^{-2}$ leads to a ratio of 
$\approx5$ between the observed \hbox{6--8} and \hbox{4--6 keV} energy output
(represented by the $EF_E$ values).\label{footnote:mytorus}}
we would expect typical column densities of $N_{\rm H}\approx
4\times10^{24}$~cm$^{-2}$.
Due to the Compton-thick nature of the sources on average, and the
fact that there are surely many star-forming galaxies in
Sample D, the stacked \xray\ spectrum (with a measured $\Gamma_{\rm eff}=1.60\pm0.16$
that is derived from the
stacked band ratio, $0.48\pm 0.08$; see Table~\ref{tab:samples}) 
will be affected by both Compton-thick AGN emission and 
star-formation emission.
The quality of the stacked \xray\ spectrum does not allow for a proper fit,
but an illustrative fit is sufficiently useful for our purposes here,
which is shown as the solid curve in Fig.~\ref{fig:spectrum}.
This is the sum of three components (each evaluated at $z=1.6$):
an unabsorbed power-law representing the star-formation
component (dotted line; the {\sc powerlaw} model in XSPEC\footnote{XSPEC is an \xray\
spectral fitting package (Arnaud 1996) that is available at http://heasarc.nasa.gov/xanadu/xspec/.} 
with $\Gamma=2.0$),
a Compton-reflection component from the AGN (dashed curve),
and a transmission component from the AGN (dashed-dot curve).
The latter two AGN components (reflection and transmission)
were obtained with a MYTORUS model with $N_{\rm H}=4\times10^{24}$~cm$^{-2}$
(see Footnote~\ref{footnote:mytorus} for the values of other parameters). 
It is clear that 
(1) the reflection component dominates the \hbox{6--8 keV} emission
(cf. a composite, reflection-dominated spectrum for a sample of highly obscured AGNs at $z\approx2$ 
obtained by Alexander et~al. 2011; see their Fig.~5),
being a factor of $\approx4$ larger than the transmission component; and 
(2) the unabsorbed power-law component dominates the \hbox{0.5--4 keV} emission.

\subsubsection{Constraints upon AGN fraction}\label{sec:con_frac}

A quantity of interest is the AGN fraction ($f_{\rm AGN}$) in a parent sample of galaxies.
While the determination of $f_{\rm AGN}$ is challenging,
there have been some previous attempts for, e.g., samples of galaxies that include \xray-detected AGNs.
For example, X10 estimated $f_{\rm AGN}\approx10\%$ for moderate-luminosity 
($L_{\rm 0.5-8\,keV}\approx10^{41.9-43.7}$ \lum) AGNs
at \hbox{$z\approx 0$--3} in a parent sample of galaxies with
$M_\star\ge 10^{10.3}\:M_\odot$.
Recently Aird et~al. (2012) studied a sample of \hbox{2--10~keV} selected AGNs and
their parent sample of galaxies that have $0.2<z<1.0$ and 
\hbox{$3\times10^9<M_\star/M_\odot<10^{12}$}.
They found that the incidence of AGN can be defined by a universal Eddington-ratio distribution
that is independent of the host-galaxy stellar mass and has a power-law form with 
the slope being $-0.65$ and the normalization evolving strongly with redshift [$\propto(1+z)^{3.8}$].
Their results, if applicable down to lower mass galaxies and for AGNs 
up to higher redshifts, 
would yield an estimate of $f_{\rm AGN}\approx10$\% for AGNs of 
$L_{\rm 2-10\,keV}\approx10^{41-44}$ \lum\ 
in galaxies with $M_\star\approx10^9\:M_{\odot}$ at \hbox{$z\approx1$--3}.

The above estimates of $f_{\rm AGN}\approx10\%$ appear to satisfy the \hbox{6--8 keV}
non-detection requirement of individual hidden AGNs in Sample~D.
The 6845 Sample~D sources have a total stacked \hbox{6--8~keV} flux of $3.5\times 10^{-14}$ \flux.
Assuming this total flux is produced uniformly from a fraction $f_{\rm AGN}$
of the Sample~D sources
(corresponding to an observed sky density of $\approx 2.5\times10^5f_{\rm AGN}$~deg$^{-2}$
given a stacking area of 0.027~deg$^2$; see \S~\ref{sec:prop}),
we obtain an average \hbox{6--8~keV} flux, $5.1\times10^{-18}f_{\rm AGN}^{-1}$~\flux,
for the hidden AGNs in Sample~D.
The on-axis \hbox{6--8 keV} sensitivity limit in the 4~Ms \cdfs\ is $\approx 2\times10^{-16}$~\flux.
If it is assumed that these hidden sources are just below the \hbox{6--8 keV} detection threshold,
then the non-detection in this band requires $5.1\times10^{-18}f_{\rm AGN}^{-1}<2\times10^{-16}$,
i.e., $f_{\rm AGN}>2.6\%$, indicating $>170$~AGNs in Sample~D.
Another estimate of a lower limit on $f_{\rm AGN}$ can be obtained through
population-synthesis models.
For instance, the Gilli et~al.~(2007) model
predicts that there are \hbox{$\approx150$}
obscured (i.e., \hbox{$N_{\rm H}\gsim 10^{22}$~cm$^{-2}$}) AGNs
with \hbox{0.5--2 keV} rest-frame intrinsic luminosities greater than $10^{42}$ \lum\
not detected in the central 6\arcmin-radius area of the 4~Ms \cdfs,
and that \hbox{$\approx 30$--50\%} of these missing AGNs are highly obscured
(i.e., \hbox{$N_{\rm H}\gsim 3\times10^{23}$~cm$^{-2}$}).
This predicted fraction of obscured AGNs, $\approx 150/6845\approx 2.2\%$ for Sample~D,
is likely a lower limit
since the population-synthesis model of Gilli et~al.~(2007) does not take into account
low-luminosity AGNs (i.e., $L_{\rm 0.5-2\,keV}<10^{42}$~\lum) that tend to be hosted by
low-mass galaxies.
Therefore, the number of missing AGNs in Sample~D
could potentially be up to several hundred.

\begin{figure}[t]
\center
\includegraphics[angle=0,scale=.77]{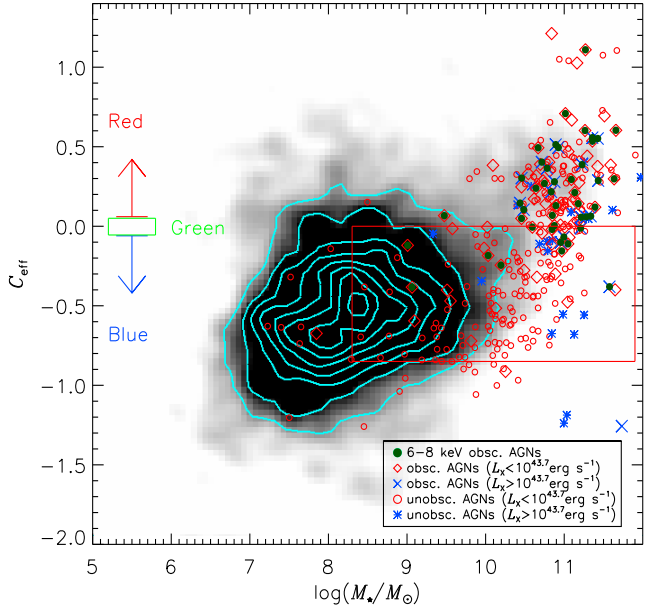}
\caption{Same as Fig.~\ref{fig:cmd}a, but including the unobscured (i.e., having $\Gamma_{\rm eff}>1$)
and obscured (i.e., having $\Gamma_{\rm eff}<1$) AGNs 
in the central $6\arcmin$~area of the 4~Ms \cdfs.
A small fraction of the AGNs are luminous (i.e., $L_{\rm 0.5-8\,keV}>10^{43.7}$~\lum),
so the color and stellar-mass estimates of their hosts are subject to AGN contamination;
however, this does not affect our discussion here (see text).}
\label{fig:cmd-unobs}
\end{figure}

In addition to $f_{\rm AGN}$, another quantity of interest is the obscured AGN fraction.
As discussed in \S~\ref{sec:spec_cons}, the majority of the underlying AGNs
in Sample~D need to be highly obscured 
in order to produce the steep \hbox{6--8 keV} rise in the stacked Sample~D spectrum;
this result suggests that the obscured AGN fraction should be close to $f_{\rm AGN}$ for 
Sample~D.
We show in Fig.~\ref{fig:cmd-unobs} the effective color-mass diagram for the
unobscured (i.e., having $\Gamma_{\rm eff}>1$)
and obscured (i.e., having $\Gamma_{\rm eff}<1$) AGNs detected
in the central $6\arcmin$~area of the 4~Ms \cdfs.
It seems clear from the figure that
the fraction of \xray\ detected sources decreases toward
lower masses regardless of whether the \xray\ sources are obscured or unobscured.
In the stellar-mass range
($2\times10^8\lsim M_\star/M_\odot\lsim 2\times 10^9$)
where most of the stacked \hbox{6--8 keV} signal lies,
there are only about 10 \xray\ detected unobscured AGNs
(and only 3 of them have $1\lsim z\lsim 3$).
An order-of-magnitude estimate of the percentage of highly obscured AGNs
among the underlying AGN population in Sample~D would then be $1-[3/(6845\times f_{\rm AGN})]>90\%$. 
This percentage of highly obscured AGNs appears high when compared to available
attempts at measurement of this quantity as a function of \xray\ luminosity and
redshift (e.g., Treister \& Urry 2006; Hasinger et~al. 2008; Gilli et~al. 2010);
see \S~\ref{sec:bh} for estimation of the typical \xray\ luminosities of our sources.
However, the available attempted measurements have significant systematic
uncertainties owing to selection incompleteness, limited source spectral
characterization, and other issues. 
Furthermore, as is clear from Fig.~\ref{fig:cmd}, we
are investigating distant AGN activity in a quite different regime from 
that where the luminosity and redshift dependences of the obscured percentage
have been studied, so past results may not be applicable.

\subsubsection{Constraints upon \xray\ luminosity, black-hole mass, and host stellar mass}\label{sec:bh}

As estimated earlier, we obtain an average \hbox{6--8~keV} flux of
$5.1\times10^{-18}f_{\rm AGN}^{-1}$~\flux\
for the hidden AGNs in Sample~D.
Adopting a reasonable absorption correction for 
a column density of $N_{\rm H}=4\times10^{24}$~cm$^{-2}$ 
within the MYTORUS model (see Footnote~\ref{footnote:mytorus} for the adopted model parameters),
we estimate the average \hbox{2--10~keV} rest-frame intrinsic luminosity 
to be $4.0\times10^{42}f_{\rm AGN}^{-1}$~\lum\ (assuming $z=1.6$).
Given the corresponding average \hbox{2--10~keV} rest-frame 
absorbed luminosity $2.0\times10^{41}f_{\rm AGN}^{-1}$~\lum,
the ratio between the absorbed and intrinsic \hbox{2--10~keV} luminosities is therefore
$(2.0\times10^{41})/(4.0\times10^{42})\approx 5.0\%$.
This ratio for Compton-thick AGNs is strongly dependent upon the
precise measurement of absorption, which is difficult and thus
renders this ratio uncertain (see, e.g., Comastri 2004 for a review).

Taking $f_{\rm AGN}=10\%$ and assuming a \hbox{2--10 keV} bolometric correction of 10
(e.g., Vasudevan et~al. 2009, 2010; Lusso et~al. 2011),
we estimate the average bolometric luminosity of the highly obscured AGNs 
hidden in Sample~D to be $4.0\times10^{44}$~\lum,
which implies that the masses of the relevant SMBH are $>3.1\times10^6\:M_\odot$
if they accrete at a sub-Eddington level. 
We assume $M_\star \sim M_{\rm bulge}$ for simplicity
and obtain a correlation of $M_{\rm BH}/M_\star\approx 1/500$ in the local universe
based on the results of Marconi \& Hunt (2003).
There are studies indicating that the average SMBH to host-galaxy mass ratio
evolves positively with redshift (e.g., Woo et~al. 2008; Merloni et~al. 2010);
however, such studies are subject to large uncertainties (e.g., Jahnke et~al. 2009;
Shen \& Kelly 2010).
Here we take an evolution form of $M_{\rm BH}/M_\star\propto (1+z)^{0.68}$ (Merloni et~al. 2010)
and obtain a correlation of 
$M_{\rm BH}/M_\star\approx (1/500)\times(1+1.6)^{0.68}\approx 1/250$ at $z=1.6$, 
which then implies a typical stellar mass of $M_\star>7.8\times 10^8\:M_\odot$
for the hosts of the highly obscured AGNs hidden in Sample~D.
This $M_\star>7.8\times 10^8\:M_\odot$ constraint is just consistent with
the median stellar mass of Sample~D ($8.1\times 10^8\:M_\odot$; see Table~\ref{tab:samples}).
However, there will be a mis-match between the estimated SMBH mass
and the typical host stellar mass,
if a lower $f_{\rm AGN}$ value or a lower ratio between 
the absorbed and intrinsic \hbox{2--10 keV} luminosities is assumed.

\subsubsection{Constraints upon star-forming galaxies}\label{sec:con_sf}

As shown in Fig.~\ref{fig:spectrum},  
the \hbox{0.5--2 keV} stacked \xray\ emission of Sample~D appears to be dominated 
by an unabsorbed power-law component that is likely associated with star-formation
related processes.
The Sample~D sources have a total \hbox{0.5--2~keV} flux of $9.8\times 10^{-15}$ \flux;
the corresponding average \hbox{0.5--2~keV} flux, $1.43\times10^{-18}$~\flux,
is a factor of $\approx6$ below the on-axis \hbox{0.5--2~keV} 
sensitivity limit in the 4~Ms \cdfs\ (X11).
We estimated absorption-corrected factors in the \hbox{0.5--2}, \hbox{0.5--8}, and \hbox{2--10~keV}
bands using a {\sc zpowerlw}$\times${\sc zwabs}$\times${\sc wabs} model in XSPEC, where
$z=1.6$, $\Gamma=2.0$, and intrinsic $N_{\rm H}=10^{20}$~cm$^{-2}$.
We then estimated the average \hbox{0.5--2}, \hbox{0.5--8}, and \hbox{2--10~keV} rest-frame intrinsic
luminosities to be $2.6\times10^{40}$, $5.2\times10^{40}$, and $3.0\times10^{40}$~\lum, respectively.
Using these luminosities, we obtained SFR estimates of 
5.7~$M_\odot$~yr$^{-1}$ based on the Ranalli et~al. (2003) relation between SFR and $L_{\rm 0.5-2\; keV}$ (see their Eq.~14),
21.8~$M_\odot$~yr$^{-1}$ based on the Lehmer et~al. (2010) relation between SFR and $L_{\rm 2-10\; keV}$ (see the fourth line in their Table~4),
19.9~$M_\odot$~yr$^{-1}$ based on the Mineo et~al. (2012) relation between SFR and $L_{\rm 0.5-8\; keV}$ (see their Eq.~22), and
4.2~$M_\odot$~yr$^{-1}$ based on the Vattakunnel et~al. (2012) relation between SFR and $L_{\rm 2-10\; keV}$ (see their Eq.~6), respectively,
assuming that the mentioned relations also apply at lower stellar masses and higher redshifts.
Based on the SFRs derived above and the mean stellar mass of the Sample~D sources
($3.2\times10^9\:M_{\odot}$),
we estimate specific SFRs (sSFR, i.e., SFR per stellar mass) ranging 
from $1.3\: \mbox{Gyr}^{-1}$ to $6.8\: \mbox{Gyr}^{-1}$.
These estimated sSFRs are on the same order of magnitude as an estimate of $2.0 \ \mbox{Gyr}^{-1}$
made using Eq.~1 of Peng et~al.~(2010) that was obtained based on observations of 
typical star-forming galaxies at \hbox{$z\approx0$--2}, 
with the input values of $3.2\times10^9\:M_{\odot}$ and $z=1.6$.
We note that the above relations between SFR and \xray\ luminosities are subject to
large uncertainties, with typical scatters of \hbox{0.4--0.5 dex}.

\subsection{Supporting Evidence for Relatively Low-Mass Galaxies Hosting Highly Obscured AGNs}\label{sec:evidence}

Our finding, that there is an appreciable fraction of
relatively low-mass galaxies that host highly obscured AGNs at \hbox{$z\approx 1$--3}, 
is somewhat unexpected.
Nevertheless, there is already some supporting evidence,
i.e., there are potential analogs both in the distant and nearby universe.
For example, Trump et~al. (2011) identified apparent weak and/or obscured AGN activity
in a sample of 28 \xray\ undetected, low-mass ($M_{\star, \rm median}\approx 3\times 10^9\: M_\odot$),
$z\approx 2$ emission-line galaxies in the \hbox{GOODS-S} region,\footnote{Of the Trump et~al.
(2011) sample of 28 galaxies, 26 (14) are included in our Sample~A (Sample~D)
and the other 2 are not included due to their vicinity to the 4~Ms \cdfs\ sources. 
Of these 26 (14) common Sample~A (Sample~D) sources, 
our estimates of $z_{\rm phot}$ and stellar mass for 18 (11) sources
are in reasonable agreement with the Trump et~al. (2011) estimates.} 
suggesting that AGNs may be common in relatively low-mass
star-forming galaxies at $z\approx2$.
Further near-infrared spectroscopic observations are needed to identify 
larger samples of highly obscured AGNs in relatively low-mass galaxies at high redshifts.
However, as demonstrated by Goulding \& Alexander (2009) and Goulding et~al. (2010),
even in the nearby universe significant mass accretion onto SMBHs could be missed
in the most sensitive optical surveys due to absent or weak optical AGN signatures
caused by extinction.  
Locally, a recent study revealed a \chandra-detected, moderately obscured
($N_{\rm H}\approx6\times10^{22}$~cm$^{-2}$) AGN that may have
$M_{\rm BH}\sim 2\times10^6\ M_\odot$ residing in a dwarf galaxy (\hbox{Henize 2-10}) with  
$M_\star\approx 3.7\times10^9\ M_\odot$ (Reines et~al.~2011).

\subsection{Future Prospects}\label{sec:prospect}

As discussed earlier, 
there are likely at least several hundred highly obscured AGNs hidden in Sample~D.
If we could better isolate this population of missing AGNs,
we would be able to boost significantly the \hbox{6--8 keV} stacked signal.
One possibility for achieving a better stacked signal
would be to obtain improved photometry that extends to the key rest-frame $K$-band or beyond.
Such improved photometry will lead to more reliable stellar-mass estimates,
which will consequently result in a more efficient sample selection and
a likely boost in the stacked signal.
The Mid-InfraRed Instrument (MIRI; Swinyard et~al.~2004; Wright
et~al.~2004) onboard the James Webb Space Telescope ({\it JWST}; Gardner et~al. 2006) will be
able to provide near- and mid-IR data that are greatly superior to the
{\it Spitzer} IRAC data currently in use. The limiting magnitudes of MIRI (for the same
length of exposure and the same signal-to-noise ratio) go deeper by over
2~magnitudes than those of IRAC. 
This means that all of our Sample~A sources
(compared to $\approx$60\% currently) will have rest-frame $K$-band
coverage or beyond with photometric quality significantly better than that at present. 
Other possibilities for increasing the stacked signal by, for instance,
identifying AGN candidates through morphologies or
through deep optical and near-infrared spectroscopic observations 
(where the CANDELS imaging and {\it JWST} spectroscopic
data would be most critical),
remain of interest, but are beyond the scope of this work.

One might think that future hard \xray\ missions such as {\it NuSTAR} and \hbox{\it ASTRO-H}
would be able to detect such highly obscured AGNs hidden in relatively low-mass galaxies.
In the distant universe ($z\gsim0.5$), however, {\it NuSTAR} and \hbox{\it ASTRO-H}
simply do not have sufficient sensitivity to make such direct detections (e.g., Luo et~al. 2011).
In contrast, a 10~Ms \cdfs\ has the potential of detecting a fraction of such highly
obscured AGNs if at least some of these hidden sources are not too far below the \hbox{6--8 keV} detection threshold
of the 4~Ms \cdfs.
Furthermore, a 10~Ms \cdfs\ would also increase the signal-to-noise ratios of the stacking results
and thus complement the aforementioned approaches of improved sample selection.

\acknowledgments
We thank the referee for helpful feedback that improved this work.
We thank T.~Dahlen for providing the \hbox{GOODS-S} {\it HST} $z$-band selected photometric catalog,
J.~R.~Mullaney and M.~Pannella for making comparisons between stellar-mass estimates, and
R.~Ciardullo and C.~Gronwall for helpful discussions.
Support for this work was provided by NASA through \chandra\ Award
SP1-12007A (YQX, SXW, WNB) issued by the \chandra\ X-ray Observatory
Center, which is operated by the Smithsonian Astrophysical
Observatory, and by NASA ADP grant NNX10AC99G (YQX, SXW, WNB).
We also acknowledge the financial support of 
the Youth 1000 Plan (QingNianQianRen) program and the USTC startup funding (YQX),
the Science and Technology Facilities Council (DMA),
\chandra\ Award SP1-12007B (FEB), the Programa de Financiamiento Basal (FEB), the CONICYT-Chile grants FONDECYT 1101024 and  FONDAP-CATA 15010003 (FEB), 
the Italian Space Agency (ASI) under the ASI-INAF contract I/009/10/0 (AC, RG, CV), and
the Einstein Fellowship Program (BDL).
\\


\end{document}